\documentclass[manuscript,screen,nonacm]{acmart}
\usepackage{tikz}
\usetikzlibrary{external}
\usepackage{cleveref,paralist}

\usepackage[breakable]{tcolorbox}
\usepackage{framed}
\usepackage{multirow}
\newtcolorbox{cotransexample}{%
    breakable,
    colback=white,
    colframe=black,
    coltitle=white,
    title={\bfseries Example 3: Co-transformation technique}
}

\newtcolorbox{taylorexample}{%
    breakable,
    colback=white,
    colframe=black,
    coltitle=white,
    title={\bfseries Example 1: First-order Taylor approximation}
}

\newtcolorbox{errorcorrectionexample}{%
    breakable,
    colback=white,
    colframe=black,
    coltitle=white,
    title={\bfseries Example 2: Error-correction technique}
}

\newtcolorbox{theorembox}{%
    breakable,
    colback=white,
    colframe=black,
    coltitle=white,
}

\usepackage{pgfplots}
\pgfplotsset{width=10cm,compat=1.9}

\usepgfplotslibrary{external}

\AtBeginDocument{%
  \providecommand\BibTeX{{%
    \normalfont B\kern-0.5em{\scshape i\kern-0.25em b}\kern-0.8em\TeX}}}

\setcopyright{acmcopyright}
\copyrightyear{2018}
\acmYear{2018}
\acmDOI{XXXXXXX.XXXXXXX}

\usepackage{color,xspace,listings,comment,paralist,tabulary}
\usepackage{algorithm,algpseudocode}
\usepackage[colorinlistoftodos]{todonotes}

\begin{document}

\title{Rigorous Error Analysis for Logarithmic Number Systems}
\author{Thanh Son Nguyen}
\affiliation{
\institution{Kahlert School of Computing - The University of Utah}
\state{Utah}
\country{USA}
}
\author{Alexey Solovyev}
\affiliation{
\institution{Kahlert School of Computing - The University of Utah}
\state{Utah}
\country{USA}
}
\author{Ganesh Gopalakrishnan}
\affiliation{
\institution{Kahlert School of Computing - The University of Utah}
\state{Utah}
\country{USA}
}

\begin{abstract}
Logarithmic Number Systems (LNS) hold considerable promise in helping
reduce the number of bits needed to represent a high dynamic range of
real-numbers with finite precision, and also efficiently support
multiplication and division.  However, under LNS, addition and
subtraction turn into non-linear functions that must be
approximated---typically using precomputed table-based functions.
Additionally, multiple layers of error correction are typically needed
to improve result accuracy.  Unfortunately, previous efforts have not
characterized the resulting error bound.  We provide the first
rigorous analysis of LNS, covering detailed techniques such as {\em
co-transformation} that are crucial to implement subtraction with
reasonable accuracy.  We provide theorems capturing the error due to
table interpolations, the finite precision of pre-computed values in
the tables, and the error introduced by fix-point multiplications
involved in LNS implementations. We empirically validate our analysis
using a Python implementation, showing that our analytical bounds are
tight, and that our testing campaign generates inputs diverse-enough
to almost match (but not exceed) the analytical bounds.  We close with
discussions on how to adapt our analysis to LNS systems with different
bases and also discuss many pragmatic ramifications of our work in the
broader arena of scientific computing and machine learning.
\end{abstract}






 
 
\maketitle

\tableofcontents
 
\section{Introduction}
\label{sec:intro}

With the increasing costs of data movement in today's 
  HPC and ML applications~\cite{data-movement-shalf,cloudy-uncertain-reed},
 there is significant pressure to reduce the number of bits used to represent real numbers using finite-precision representations.
With fewer bits moved, memory bandwidth as well as cache memories are better utilized.
Also, given the sheer number of
scalar multiplications 
carried
out by 
these applications (e.g., when performing 
matrix and tensor products), those
number representations that help  reduce multiplication (and division) costs are also of great importance.
Logarithmic number systems (LNS)  possess both these advantages.
They store only the logarithm of real numbers in 
{\em finite-precision fixed-point representations}\footnote{In virtually all LNS implementations that are surveyed later.}.
Furthermore, multiplications and divisions turn into fixed-point addition and fixed-point subtraction (respectively); and in the absence of overflows, are exact.
Also, square-root turns into division by 2 (right shift), thus exact.
%

Addition and subtraction are a whole different story, turning into calculations involving
non-linear functions.
This requires good trade-offs between error control and computational speed.
In more detail, methods are needed to perform additions and subtractions with techniques to perform {\em multiple levels of error correction} using various lookup tables whose overall cost must be minimized.\footnote{``Tableless methods'' are also popular, approximating what table-based methods provide, but are not studied in this work.}

However, despite the availability of these error correction methods, {\em there are no rigorous error analysis methods available that tightly bound the worst-case error.}
This is quite an odd situation: on one hand, we have methods to compensate for errors.
On the other hand, no one has produced a tight bound on the worst-case error.
Our work in this paper closes this gap by providing such a rigorous error bound.

All existing error estimation techniques that we know of perform error estimations through empirical testing.
It is well-known that without tight error bounds, actual hardware/software designs most likely will end up over-provisioning to accomodate for these ``excess errors'' that never occur.
The key contribution of this paper is the first such tight and rigorous bound parameterized over today's popular LNS schemes.
Specifically,
\begin{compactitem}
    \item We provide the first tight parametric error estimate formulae in terms of parameters such as 
    the 
     machine-epsilon of
     fixed-point numbers as well as
     look-up table sizes.
Such a parameterization
can help precisely guide
hardware and software implementations.

    \item We demonstrate through systematic testing using a Python implementation of LNS  that our hand-derived estimates
    are trustworthy {\em and tight}. We release this code
    to enable others to reproduce and further validate our work.
\end{compactitem}
A more rigorous mechanical verification is not attempted at this stage for the 
following reasons:
\begin{compactitem}
    \item None of the existing SMT-based~\cite{cvc5,z3,yices} automated
    tools are sufficiently
    powerful to check the validity our
    derivations.
    
    \item Our analytical results and the manner
    in which
    we decompose their proofs are 
    similar to those
    necessary in any
    attempt
    at mechanical
    proofs 
    using expert-guided proof-checkers---efforts that are typically multi-year, and beyond the scope of this paper.
\end{compactitem}

\paragraph{Comparison of LNS and IEEE Floating-Point}
For those familiar with IEEE floating-point arithmetic~\cite{muller2018handbook}, 
an IEEE 
{\em normal}
floating-point number is described by a triple $(s,e,m)$, where $s\in \{-1,1\}$ denotes the sign,
$e \in \mathbb{Z}$ is the exponent and $m \in [1,2)$ is the mantissa (also known as the significant).
The value of the real number represented by this triple is 
$s\cdot (2^e)\cdot m$.
Unlike floating-point, LNS does not have the mantissa but allows rational exponent instead.
Also, while a {\em subnormal} IEEE floating-point number has $m < 1$ (in which case its value is encoded
by the value in the mantissa weighted by the smallest normal
exponent),
LNS does not have the notion of subnormals: the entire representable number scale is modeled in the same manner.
Specifically, an LNS number is described by a pair $(s,e)$, where $s\in \{-1,1\}$ denotes the sign and
$e\in \mathbb{Q}$ is the exponent.
This pair represents the real value $s\cdot 2^e$.
The exponent part of LNS is represented by a signed fixed-point number
with a special arrangement to represent $0$.

\paragraph{Rounding modes for LNS}
Analogous to IEEE floating-point, two consecutive
LNS fixed-point words define an interval of real-numbered values, and any real-number falling in
this interval may be rounded up, down, or to
the nearest value.
The upper-bound of this distance of rounding
(depending on the rounding mode) defines the
{\em machine-epsilon} of LNS, which we denote
by $\epsilon$ in the rest of this paper.

\begin{figure}[htb]
\includegraphics[scale=1]{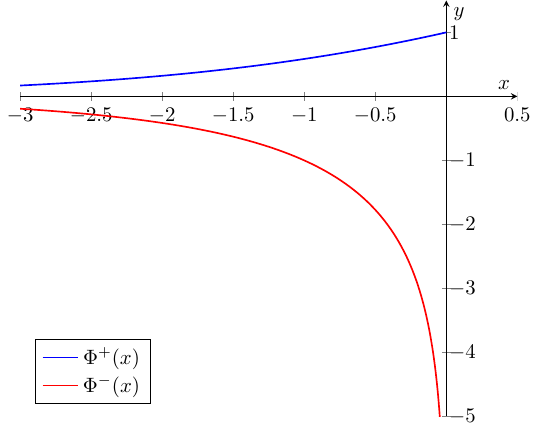}
  \caption{Plots of $\Phi^+(x)$ and $\Phi^-(x)$}
  \label{fig:phi}
\end{figure}

\paragraph{Rigorous Problem Statement}
We now introduce some basic notions underlying LNS that allow us to define the problem a bit more tightly.
First, we describe how the four basic operations: addition, subtraction, multiplication
and division are performed in LNS (operations such as square-root are not described, for brevity).
In all four operations, the sign and magnitude of the result can be computed separately.
Because computing the sign is straightforward, 
to make it easy to read, we assume that all the operands are positive
to demonstrate how LNS computes the magnitude of the result.
Let $2^p, 2^q$ be real numbers which can be exactly represented in LNS by
the two fixed-point numbers $p,q$, then multiplication and division of $2^p$ and $2^q$
can be performed efficiently and exactly in LNS by fixed-point addition and subtraction:
$$ \log_2(2^p\times 2^q) = \log_2 2^{p+q} = p+q $$
$$ \log_2(2^p/2^q) = \log_2 2^{p-q} = p-q $$
However, the main drawback is that addition and subtraction are not directly realizable in LNS. 
Without loss of generality, let $p \geq q$ and 
let us use $x\leq 0$ to denote $q-p$ (this allows us to write $2^x$, knowing that it will be a fraction in $(0,1]$).
Then, 
$$\log_2(2^p + 2^q) = \log_2 (2^p (1+2^{x}))= i+ \log_2(1+2^{x}) $$   
$$\log_2(2^p - 2^q) = \log_2 (2^p (1-2^{x}))= i+ \log_2(1-2^{x})$$
%
Now let us introduce two non-linear functions
$\Phi^+$
 and 
 $\Phi^-$
 (plotted in
 Figure~\ref{fig:phi}, and also called
 {\em Gaussian Logarithm}~\cite{gaussianlog}) defined as:
$$\Phi^+(x) = \log_2(1+2^x) \quad \text{and} \quad \Phi^-(x) = \log_2(1-2^x)$$
In Section~\ref{sec:background},
we will show that 
 $\Phi^+$ and  $\Phi^-$
will be approximated via ROM tables look-up and interpolation.
Thus, to compute 
$\log_2(2^p + 2^q)$,
 we can simply perform a {\em fixed-point addition
 of} $p$
 to
 the result of
$\Phi^+(x)$
(likewise for 
$\log_2(2^p - 2^q)$).

{\bf In this paper, we focus on deriving the bounds for the absolute-errors of the
approximations of  $\Phi^+$ and  $\Phi^-$, for convenience, we call them {\em error-bounds }
by default.}

%
%
\paragraph{Rigorous error-bound derivation}
We derived the error-bound of the three most popular 
$\Phi^+$/$\Phi^-$ approximation techniques 
which are proposed in European Logarithm Microprocessor~\cite{AEM}.
%
%
%
These techniques will be described in detail in Section~\ref{sec:background}.
To derive the rigorous error-bounds,
firstly, we analyze all error sources of each technique,
which include the errors of the mathematical nature of
the approximation method and the errors of finite-precision 
hardware implementation.
Then, we mathematically derive the error-bound of each
error source by symbolic function analysis, 
and finally accumulate them using the absolute-norm inequality.
The mathematics derivations are performed manually with
significant support from the Sympy~\cite{sympy} library,
which handles symbolic expression manipulations.
Details of all error sources considered in our analysis are mentioned in Section~\ref{sec:high-level-overview-math}
and details of 
error-bounds derivations
for each source
are presented in Sections~\ref{sec:Taylor},~\ref{sec:EC} and ~\ref{sec:cotrans}.

\paragraph{Validation of Rigorous Analysis: Key Results}
In our validation of
our bound tightness,
we compute the values of $\Phi^+$ and $\Phi^-$
over large and dense input sample sets, and across different LNS configurations.
We then compare the results against the results
of direct computation of our analytical formulae---all using extended precision.
From our experimental results, the fact that none of the errors exceeds the error-bounds
computed using our error-bound formulae.
We also control our test-case generation process to 
produce
instances where
the error is very close
to (but never exceeds) the error-bounds.
We provide detailed histograms 
of inputs showing that
the concentration of inputs
where these inputs are low,
but non-zero---all this
going to support
the correctness and tightness of our derived error-bound formulae.

\paragraph{Organization} In Section~\ref{sec:relwork}, 
we survey  existing LNS approaches that we are aware of.
Section~\ref{sec:background} introduces some background and the notations used in our derivations.
Because of the tedious and detailed nature of our error analysis, in Section~\ref{sec:high-level-overview-math},
we will present all of our derivations via plots and intuitive explanations.
Then, Section~\ref{sec:Taylor} contains the details of
error-bound derivation of first-order Taylor approximation (can be skipped upon first reading).
Also, Section~\ref{sec:EC} 
and Section~\ref{sec:cotrans} contain the details of error-bound derivation
of the error-correction technique and the co-transformation technique (also can be skipped upon first reading).
Section~\ref{sec:numerical-expts} checks our analysis through numerical experiments.
Section~\ref{sec:conc} describes adaptations of the LNS we analyze to other LNS---showing the possibility of easily carrying over our approach to these other LNS without deriving their error-bounds from scratch, then our conclusions and future work.

\noindent{\bf Code Release:\/}
We publish our github repository at \url{https://github.com/Thanhson89/RigorousErrorLNS}, 
which includes the tool for calculating LNS's rigorous error-bound with respect to the user's
input parameters and the result of our experiments
in Section~\ref{sec:numerical-expts}, and details 
of symbolic algebra tools used.

\section{Related Work}
\label{sec:relwork}

%







Proposed in the early 1970s,
LNS are still a topic
of current active
interest.
Many of these proposed  schemes 
have been realized in software.
An LNS-based implementation of weight updates in neural network training was recently proposed by NVIDIA~\cite{madam} where a hardware implementation is proposed.
A recent hardware implementation of the Ising model~\cite{ang-li-lns-dac-2023}
also employs the LNS.
A bibliograph of over 600 citations relevant to LNS research has been provided by XLNS Research~\cite{xlns-research}.

Unfortunately, {\em no prior work has derived rigorous  error bounds}, which is our main contribution here.
We now briefly survey prior LNS, citing the logarithm-base used in them, how addition and subtraction are realized, whether a hardware implementation exists, and the status of error analysis to the extent we are aware of.
%


\paragraph{European Logarithm Microprocessor~\cite{EM}}
This is a historically 
important
piece of work resulting in
the first
complete hardware realization of an LNS microprocessor.
This LNS scheme
        uses  base-$2$
        logarithm.
        It employs the
         error-correction algorithms for the
         approximation of $\Phi^\pm(x) = \log(1\pm 2^x)$,
         and also
         employs
         the co-tranformation technique~\cite{cotrans} for $\Phi^-$ when $x \to 0$
(both these ideas are detailed in
Section~\ref{sec:pd}).  %
These authors evaluate the area and delay of their LNS implementation, showing that while the area for the LNS realization is equivalent to that of floating-point implementations, the delay could be much better.
    Although the authors compare the error of
    LNS with that of floating-point through empirical testing, 
    there is no rigorous error analysis for LNS.
     
     \paragraph{Low-precision LNS beyond Base-2~\cite{LNSbeyond2}}
    In this LNS scheme, the logarithm base is
        in the range $\sqrt{2}$
        to $2$.
        The base is
        selected so as to optimize the approximation error of the
        inputs as well as 
        the sizes of addition/subtraction truth tables.
        This paper also suggests that it is more efficient to 
        implement these tables using logic gates rather than using ROM.
        Hardware implementations are proposed,
        and
        error analysis is
        achieved 
        via simulation for conversion and single-operators.

    \paragraph{Semi-LNS~\cite{semilns}}
This number system is a hybrid between Floating-point and LNS,
which consists of both 
of rational logarithmic and mantissa parts, for balancing the efficiency between multiplication/division and addition/subtraction.
Although there is some error analysis for representation error, that of operations' error is not mentioned.

\paragraph{ROM-less LNS} 
This LNS design is tailored for minimal hardware implementation.
The computation of addition and subtraction is based on Mitchell's method~\cite{mitchellLNS}, 
which is very efficient to be implemented on hardware with the trade-off of accuracy.
Some approaches have been proposed to improve the accuracy of Mitchell's method,
such as using operand decomposition~\cite{arnold2006}, base selection~\cite{arnold2019},
and Approximate tableless LNS ALU~\cite{arnold2020}.

\paragraph{Convolution Neural Networks (CNN) Using Logarithmic Data Representations~\cite{CNNLNS}}
In this approach, the logarithm
base is $2$ as well as $\sqrt{2}$.
 This approach
 proposes
 the first CNN implemented in low-precision LNS and showing that LNS is superior to fixed-point arithmetic in such implementation because of the non-uniform distribution of weights and activations.
 This scheme has not been implemented
 in hardware as far as we know, neither has
 rigorous analysis has been conducted.

\paragraph{LogNet~\cite{lognet}}
In this approach, the logarithm bases
used are $2$ and $\sqrt{2}$.
This work focuses on
improving the learning algorithm of CNNs
using LNS,
demonstrating the superiority of LNS over 
fixed-point schemes in terms of hardware costs.
While there is a hardware implementation,
rigorous error analysis has not
been attempted.

\paragraph{LNS-MADAM~\cite{madam}}
In this scheme, the logarithm base
used is $2^{1/k}$.
The highlight
of their work
is proposing a training algorithm for Deep Neural Networks (DNN) in low-precision 
LNS with an approximation of addition technique 
and the Multiplicative Weight Update (MWU) algorithm to replace Stochastic Gradient Descent. 
The authors propose their own LNS's design and 
hardware implementation, which is based on Mitchell's method and stochastic rounding~\cite{stochastic-rounding}.
The paper also performs symbolic error analysis just for the sake of
indicating that Multiplicative Weight Update is superior to Stochastic Gradient Descent
in terms of minimizing the quantization error-bound.
However, the error-bound is not approximated tightly enough to provide insight to
the accuracy of their LNS's design and implementation.

\paragraph{Comparison:\/} Our rigorous analysis is applicable to all these LNS variants, 
as it is parametric over the base of the logarithm, and accommodates various precision choices and table-based implementations.
\section{Background, Notations}

\label{sec:pd}
\label{sec:background}

Let us consider LNS implementations where 
multiplication and division (which
turn into 
addition and subtraction, respectively) are suitably guarded against
overflows (for example, by checking that the  underlying fixed-point addition and subtraction do not overflow or underflow). 
The rest of this paper
concentrates on addition and subtraction.

The
computation of LNS addition and subtraction involves
addition/subtraction and  interpolations
 with respect to tables
 that discretize
 the $\Phi^+$ and  $\Phi^-$ functions, suitably
 limiting the number
of pre-computed and stored  table values.
Let $\Phi$ generically stand for either $\Phi^+$ or $\Phi^-$.

When it is necessary to make it clear that
we are indexing tables, we will use the
notation $\Phi_T$.
We will sample $\Phi$
at a spacing of $\Delta$
and store these values
in $\Phi_T$.
Now,  given an arbitrary $x$,  define $\mathbf{i}$
as $\left\lceil \dfrac{x}{\Delta} \right\rceil \Delta$ and $\mathbf{r}$ as $\mathbf{i} - x$. We have $x = \mathbf{i} - \mathbf{r}$
\footnote{Strictly we must use $\mathbf{i}_x$ and $\mathbf{r}_x$
as they depend on $x$; but we prefer   
$\mathbf{i}$ and $\mathbf{r}$
for readability.}.
Recall that $x$ is negative.
Thus, $\mathbf{i}$
 is the discrete index {\em after} $x$.
It must be clear that 
$\Phi_T(\mathbf{i}) = \Phi(\mathbf{i})$
and 
$\Phi_T(x)$ (viewed as a tabular function)
is undefined at other $x$ than at
these $\mathbf{i}$.
For $x$ not in $\Phi_T$,  we can apply interpolation
techniques~\cite{EM,AEM,64bitLNS} that we will explain in 
detail, to make this paper self-contained.

\paragraph{Approximating $\Phi(x)$ through Taylor Approximation, yielding function $\hat{\Phi}_T$}



We can now define an approximation
to $\Phi(x)$ defined with the help of the two tables $\Phi_T$
and
$\Phi_T^{'}$
that are
 defined at the $i$ points:
\begin{equation}\label{eq:T}
\hat{\Phi}_T(x) = \Phi (\mathbf{i})  - \mathbf{r}\Phi'(\mathbf{i}) 
= \Phi_T (\mathbf{i})  - \mathbf{r}\Phi_T'(\mathbf{i})
\end{equation}


%


\begin{figure}
\includegraphics[scale=1]{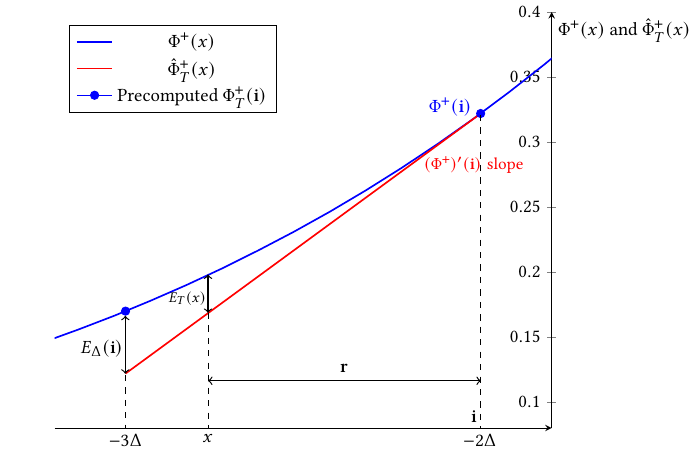}
\caption{Details of defining
$E_{\Delta}(\mathbf{i})$}
\label{fig:details_of_E_delta}
\end{figure}

\begin{taylorexample}
\label{ex:no1}
This example illustrates how $\Phi^+(x)$ is calculated using first-order Taylor approximation.
All the numerical values in this example are in binary.
Let the signed fixed-point representation of the LNS consist of 1 integer bit and 8 fractional bits.
The two look-up tables  $\Phi_T^+(x)$ and $(\Phi_T^+)'(x)$ 
are shown in Table~\ref{tab:lns-illustration} ($\Delta = 0.10000000$).
\begin{center}
\begin{tabular}{|c|c|ll|c|c|}
\cline{1-2} \cline{5-6}
\multicolumn{1}{|c|}{$\mathbf{i}$} & \multicolumn{1}{c|}{$\Phi_T^+(\mathbf{i})$} &  &  & \multicolumn{1}{c|}{$\mathbf{i}$} & \multicolumn{1}{c|}{$(\Phi_T^+)'(\mathbf{i})$} \\ \cline{1-2} \cline{5-6} 
-0.00000000                     & 1.00000000                         &  &  & -0.00000000                    & 0.10000000                            \\
-0.10000000                     & 0.11000110                         &  &  & -0.10000000                    & 0.01101010                            \\
-1.00000000                     & 0.10010110                         &  &  & -1.00000000                    & 0.01010101                         \\
-1.10000000                    & 0.01110000                         &  &  & -1.10000000                    & 0.01000011                         \\ \cline{1-2} \cline{5-6} 
\end{tabular}
\captionof{table}{Illustration of LNS via an example: two look-up tables  $\Phi_T^+$ and $(\Phi_T^+)'$ for performing First-order Taylor approximation (see Equation~\ref{eq:T}) are created}
\label{tab:lns-illustration}
\end{center}
Suppose that we want to calculate $\Phi^+(x)$ with $x = -0.11000000$:
\begin{compactenum}
\item
First, determine the index value $\mathbf{i}$
above $x$, which turns out
to be $\mathbf{i}= -0.10000000$.
This is $\mathbf{r}$
above $x$,
where
$\mathbf{r} = \mathbf{i}-x = 0.01000000$.

\item 
Next, look up 
$\Phi^+(\mathbf{i})$ and $(\Phi^+)'(\mathbf{i})$, obtaining
$\Phi^+(-0.10000000) = 0.10010110$ and $(\Phi^+)'(-0.10000000) = 0.01101010$.

\item
Finally,   $\Phi^+(x)$ using First-order Taylor approximation  (Equation~\ref{eq:T}) is
$$\hat{\Phi^+}_T(x) = \Phi_T^+(\mathbf{i}) - \mathbf{r}(\Phi_T^+)'(\mathbf{i}) $$ 
$$= 0.10010110 - 0.01000000\times 0.01101010= 0.01111011$$
\end{compactenum}
%
%
\end{taylorexample}  

\begin{figure}[h]
\includegraphics[scale=1]{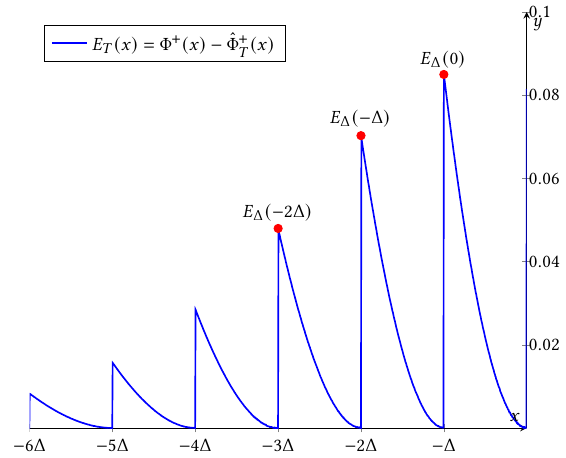}
\caption{We do first-order Taylor approximation
of $\Phi(x)$, obtaining 
$\hat{\Phi}_T$. This still has an 
error $E_T$. We plot  $E^{+}_{T}$ here.}
  \label{fig:errortaylor}
\end{figure}

\begin{figure}[h]
  \includegraphics[scale=1]{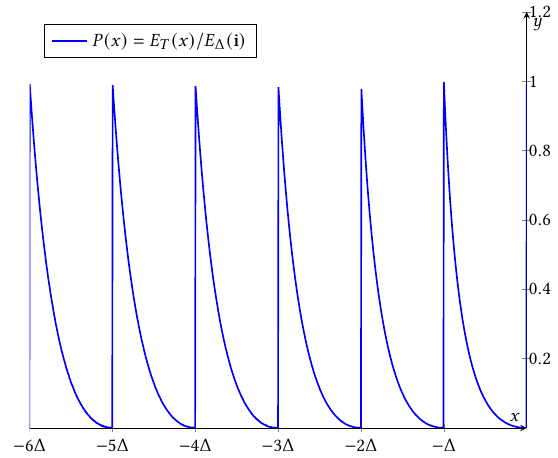}
  \caption{Error ratio $P(x)$ in each $\Delta$ interval}
  \label{fig:errorratio}
\end{figure}

For higher accuracy, the European Logarithmic Microprocessor~\cite{EM,AEM} suggests adding an
{\bf error-correction} term to the previous formula because a linear interpolation is too coarse.
Let us refer to the difference between
the mathematical
$\Phi(x)$ and 
the just now
defined approximation
$\hat{\Phi}_T(x)$ by a new
``error function''
$E_T(x)$,
which is
plotted in 
Figure~\ref{fig:errortaylor}.
A key observation made in~\cite{EM,AEM} is that 
if all the
``spikes''
of $E_T(x)$ bounded by   $\Delta-$ 
are
scaled such that
the tips of the 
spikes are $1$,
such a scaled
function, now called $P(x)$,
becomes nearly
periodic, as shown in Figure~\ref{fig:errorratio}.
This ``error-ratio function''
can be defined
as follows:
\begin{equation}
    P(x) = E_T(x)/E_{\Delta}({\bf i})
\end{equation}
where
$E_{\Delta}(\mathbf{i})$ 
is the supremum of 
$E_T$ in the
$\Delta-$segment $(\mathbf{i}-\Delta,\mathbf{i}]$
containing $x$. 
In other words, $E_{\Delta}(\mathbf{i})=  \lim_{x\to (\mathbf{i}-\Delta)} E_T(x) $.
Figure~\ref{fig:details_of_E_delta}
shows that
this can be
written as
$\Phi(\mathbf{i}-\Delta) -\Phi (\mathbf{i})  + \Delta\Phi'(\mathbf{i})$
or even as 
$\Phi_T(\mathbf{i}-\Delta) -\Phi_T (\mathbf{i})  + \Delta\Phi_T'(\mathbf{i})$

%


While function
$P(x)$
looks periodic,
strictly it is not; i.e., it
does not
not satisfy
 $P(x)=P(x-\Delta)$
but {\em it nearly does.}
We can exploit
this fact
by choosing
an arbitrary
repeating segment
of $P(x)$
and normalizing
our calculations with respect to that segment.
 More formally,
\begin{compactitem}
\item 
 Let
$k\in \mathbb{N}$ and let $c=k\Delta$.

\item Then precompute the values of $P(x)$ inside this
$c$-th segment,
filling a new
table called
$P_c$ (this table depends on the choice of $c$).

\item When we index $P_c$ at $\mathbf{r} \in [0,\Delta)$
(we will later 
make it clear what the resolution of $\mathbf{r}$ is),
we are obtaining $P(c-\mathbf{r})$, and this
corresponds to the ratio $E_T(c-\mathbf{r})/E_{\Delta}(c)$.

\item Now we can approximate $P(x)$ using the
$P_c$ table by using 
$P(x) \approx P(c+\mathbf{i}-x) = P_c(\mathbf{r})$.

\item Next, we create another lookup table
$E_{\Delta}$ that delivers the $E_T$ values
at all $\mathbf{i}$ using the ``original'' first-order
Taylor approximations.
More specifically,
$E_{\Delta}(\mathbf{i}) =
    \Phi(\mathbf{i}-\Delta) -\Phi (\mathbf{i})  + \Delta\Phi'(\mathbf{i})$.

\item The size and index values of this $E_{\Delta}$
 table are the same as those of the look-up tables for
$\Phi_T$ and $\Phi'_T$.
\end{compactitem}


With all this,
  $E_T(x)$ can be approximated by the \textbf{error-correction term}  $E_{\Delta}(\mathbf{i})P_c(\mathbf{r})$.
which is {\em added} to the first-order Taylor interpolation formula to derive a more accurate approximation (we add this correction because, as is clear from
Figure~\ref{fig:details_of_E_delta},
the original
$\hat{\Phi}_T(x)$
function consistently delivers values
smaller than the
true $\Phi(x)$
function):

\begin{equation}\label{eq:EC}
\hat{\Phi}_{EC}(x) = \hat{\Phi}_T(x) +  E_{\Delta}(\mathbf{i})P_c(\mathbf{r}) =\Phi_T (\mathbf{i})  - \mathbf{r}\Phi_T'(\mathbf{i}) + E_{\Delta}(\mathbf{i})P_c(\mathbf{r})
\end{equation}
where  
$E_{\Delta}(\mathbf{i})=  
\Phi_T(\mathbf{i}-\Delta) -\Phi_T (\mathbf{i})  + \Delta\Phi_T'(\mathbf{i})
$
and $P_c(\mathbf{r})= E_T(c-\mathbf{r}) / {E_{\Delta}(c)}$.

%
 
%
 We refer to this definition of 
 $\Phi_{EC}(x)$
 as {\bf computing
 using  the
error correction technique},
requiring 
altogether
four look-up tables: one for $E_\Delta$ 
 indexed by
 $\mathbf{i}$,
 one  for  $P_c$
 indexed by
 $\mathbf{r}$, and the former two tables 
 $\Phi_T$
 and 
 $\Phi_T^{'}$.

\begin{errorcorrectionexample}

This example illustrates how $\Phi^+(x)$ is calculated using the error-correction technique.
The fixed-point representation of the LNS and two look-up tables of $\Phi^+$ and $(\Phi^+)'$ 
are exactly the same as those of Example 1.
The two look-up tables for  $E_\Delta$ and $P_c$ are shown in  Table~\ref{tab:ec-illustration}.
\begin{center}
\begin{tabular}{|c|c|ll|c|c|}
\cline{1-2} \cline{5-6}
\multicolumn{1}{|c|}{$\mathbf{i}$} & \multicolumn{1}{c|}{$E_\Delta(\mathbf{i})$} &  &  & \multicolumn{1}{c|}{$\mathbf{r}$} & \multicolumn{1}{c|}{$P_c(\mathbf{r})$} \\ \cline{1-2} \cline{5-6} 
-0.00000000                     & 0.00000110                        &  &  & 0.00000000                    & 0.00000000                            \\
-0.10000000                     & 0.00000101                         &  &  & 0.00100000                    & 0.00001111                            \\
-1.00000000                     & 0.00000101                         &  &  & 0.01000000                    & 0.01000001                            \\
-1.10000000                     & 0.00000100                         &  &  & 0.01100000                    & 0.10101010                            \\ \cline{1-2} \cline{5-6} 
\end{tabular}
\captionof{table}{Illustration of LNS via an example: two look-up tables of $E_\Delta$ and $P_c$ for performing error-correction technique (see Equation~\ref{eq:EC})}
\label{tab:ec-illustration}
\end{center}
We want to calculate $\Phi^+(x)$ for the same value of $x = -0.11000000$ as in Example 1:
\begin{compactenum}
\item
Firstly, we calculate $\Phi^+(x)$ by First-order Taylor approximation and get 
$\hat{\Phi}^+_T(x)  =0.10000000$ (see Example 1). 
The values of $\mathbf{i} = -0.10000000$ and $\mathbf{r} = 0.01000000$ are also the same as
those in Example 1.
\item 
Secondly, we look up the values of $E_\Delta(\mathbf{i})$ and $P_c(\mathbf{r})$ in the two look-up tables and get:
$E_\Delta(-0.10000000) = 0.00000101$ and $P_c(0.01000000) = 0.01000001$. 
Note that in case $\mathbf{r}$ is not in the table $P_c$, we take the values from the closest index.
Then, we calculate the error-correction term: 
$$E_{\Delta}(\mathbf{i})P_c(\mathbf{r})= 0.00000101 \times 0.01000001 = 0.00000001$$
\item
Finally, we add the error-correction term to the First-ordered Taylor approximation (see Equation~\ref{eq:EC}) and get:
$$\hat{\Phi}^+_{EC}(x) = \hat{\Phi}^+_T(x) +  E_{\Delta}(\mathbf{i})P_c(\mathbf{r}) = 
0.01111011 + 0.00000001=0.01111100$$
\end{compactenum}
%
%
\end{errorcorrectionexample}

\paragraph{{\bf Co-transformation}:\/ Error control when $x$ approaches $0$:}
%
One of the more difficult cases of error control in LNS is when
computing $\Phi^-(x)$ for values of
$x$ is close to $0$.
The trouble arises because of the nature of this function:
the $n$th derivative
of $\Phi^-(x)$ for all $n$
tend to 
$-\infty$ as $x$ approaches $0$, i.e. $\Phi^-(x)$ has a singularity at $0$ (see Figure~\ref{fig:phi}).
To avoid computing $\Phi^-(x)$ in this range, addition/subtraction
can  be 
split across different intervals and computed by 
the so-called {\em co-transformation}
techniques~\cite{cotrans,arnoldcotrans,basirSDcotrans,basirdoublecotrans,hicotrans}.
The state-of-the-art co-transformation technique is proposed 
by European Logarithmic Microprocessor~\cite{cotrans,improvedcotrans}, 
which suggests 3-way interval split defined by
design-specific constants $\Delta_a$
and $\Delta_b$.
The general idea 
is to maintain
three extra look-up tables
$T_a, T_b$ and $T_c$
of $\Phi^-$ inside the range $(-1,0)$,
then transform $\Phi^-(x)$ such that
it can be computed by indexing those look-up tables 
together with
interpolating $\Phi^-$ outside of the range $(-1,0)$.
%
%
Specifically, the three look-up tables
$T_a, T_b$ and $T_c$, then the co-transformation technique
are described as follows  
(see Appendix~\ref{ss:proofcotrans} for the proof of correctness):

Let $\Delta_a$ and $\Delta_b$ are two positive fixed-point numbers
such that
$\Delta_a$ is very close to $0$,
$\Delta_b$ is a multiple of $\Delta_a$ and a divisor of $1$:
\begin{compactitem}
\item The table
$T_a$ covers all fixed-point numbers in a very small
range $[-\Delta_a, 0)$,
 
\item The table
$T_b$ covers all multiples of  $\Delta_a$ in the range
$[-\Delta_b, -\Delta_a)$.

\item The table
$T_c$ covers all multiples of  $\Delta_b$ in the range
$x \in (-1, -\Delta_b)$. 
\end{compactitem}
When $x$ in $(-1,0)$, the value of $x$ must belong to 
one of the three cases: 
$x \in [-\Delta_a, 0)$,
$x \in [-\Delta_b, -\Delta_a)$
and 
$x \in (-1, -\Delta_b):$.
The formulae for $\Phi^-(x)$ is derived across the three
cases as follows:
%

\vspace{1ex}
\noindent$\bullet\; ${\bf Case 1: $x \in [-\Delta_a, 0)$:\/} $\Phi^-(x)$ is indexed directly from table $T_a$.

\vspace{1ex}
\noindent$\bullet\; ${\bf Case 2: $x \in [-\Delta_b, -\Delta_a)$:\/}
\\
Let $r_b = (\left\lceil\dfrac{x}{\Delta_a}\right\rceil - 1)\Delta_a, \quad$ (i.e. $r_b$ is the index value of $T_b$ which is smaller than and closest to $x$)
\\
and $r_a = r_b - x$,
\\
and let $ k = x-\Phi^-(r_b)+\Phi^-(r_a)$.
\\
Then
\begin{equation}\label{eq:CT2}
\Phi^-(x) = \Phi^-(r_b) + \Phi^-(k)
\end{equation}
where $\Phi^-(r_a)$ and $\Phi^-(r_b)$ are indexed directly from tables $T_a$ and $T_b$ respectively, and
$\Phi^-(k)$ is computed by interpolation 
(either by first-order Taylor approximation or the error-correction technique).

\begin{figure}[h]
\includegraphics[scale=0.3]{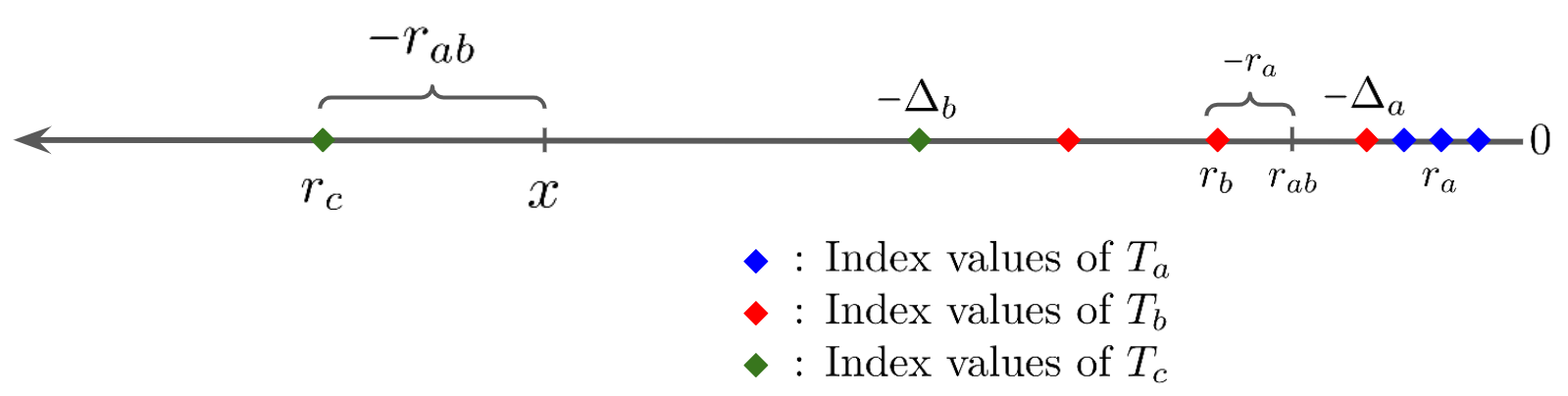}
  \caption{Co-transformation Illustration for the case where
  $x\in
  (-1,
  -\Delta_b)$:  The positions of $r_c$,$r_{ab}$,$r_b$ and $r_a$ that are derived from this $x$ are shown. Here,
        $r_c$ is the closest index value of $T_c$ on the left of $x$.
       The gap of $x$ and $r_c$ is $-r_{ab}$.
       now
       $r_b$ is the closest index value of $T_b$ on the left of $r_{ab}$.
       Finally,
      the gap of $r_{ab}$ and $r_b$ is $-r_a$.
}
  %
 \label{fig:cotranaxis}
\end{figure}

\vspace{1ex} 
\noindent$\bullet\; ${\bf Case 3: $x \in (-1, -\Delta_b)$:\/}
\\
Let 
$r_c = (\left\lceil\dfrac{x}{\Delta_b}\right\rceil -1)\Delta_b, \quad$ (i.e. $r_c$ is the index value of $T_c$ which is smaller than and closest to $x$) \\
and $r_{ab} = r_c - x$,
\\
and
$r_b = (\left\lceil\dfrac{r_{ab}}{\Delta_a}\right\rceil -1)\Delta_a, \quad$ (i.e. $r_b$ is the index value of $T_b$ which is smaller than and closest to $r_{ab}$)
\\
and $r_a = r_b - r_{ab}$,
\\
and
$k_1 = r_{ab}+\Phi^-(r_a) -\Phi^-(r_b)$,
\\
and
$k_2 = x+\Phi^-(r_b) + \Phi^-(k_1)-\Phi^-(r_c)$.
\ \\
\\
Then
\begin{equation}\label{eq:CT3}
\Phi^-(x) = \Phi^-(r_c)+\Phi^-(k_2)
\end{equation}
where $\Phi^-(r_a)$, $\Phi^-(r_b)$, and $\Phi^-(r_c)$ 
are indexed directly from 
tables $T_a$, $T_b$, and $T_c$ respectively, and
$\Phi^-(k_1)$ and $\Phi^-(k_2)$ are computed by interpolation
(either by first-order Taylor approximation or error-correction technique).
Figure~\ref{fig:cotranaxis} illustrates the positions and meanings of $r_c$, $r_{ab}$, $r_b$ and $r_a$
with respect to $x$.

\begin{cotransexample}

The following example illustrates the computation of $\Phi^-(x)$ using the co-transformation technique.
Let the signed fixed-point representation of the LNS consist of 3 integer bits and 6 fractional bits.
The co-transformation technique is applied when $x\in (-1,0)$; otherwise, we use first-order Taylor approximation.
Assume that, the two look-up tables of $\Phi^+$ and $(\Phi^+)'$ are pre-computed with $\Delta = 0.01$.
Let $ \Delta_a=0.0001$, $\Delta_b = 0.01$,
and the three tables $T_a, T_b, T_c$ are shown in Table~\ref{tab:cotrans-illustration}. 
\begin{center}
\begin{tabular}{|c|c|ll|c|c|ll|c|c|}
\cline{1-2} \cline{5-6} \cline{9-10} 
\multicolumn{2}{|c|}{$T_a$}  &  &  & \multicolumn{2}{c|}{$T_b$}  &  &  & \multicolumn{2}{c|}{$T_c$}  \\ \cline{1-2} \cline{5-6}  \cline{9-10} 

\multicolumn{1}{|c|}{$x$} & \multicolumn{1}{c|}{$\Phi^-(x)$} &  &  & \multicolumn{1}{c|}{$x$} & \multicolumn{1}{c|}{$\Phi^-(x)$} &  &  & \multicolumn{1}{c|}{$x$} & \multicolumn{1}{c|}{$\Phi^-(x)$} \\ \cline{1-2} \cline{5-6}  \cline{9-10} 
-0.000001    & -110.100010   &  &  & -0.000100     & -100.100100    &  &  & -0.010000     & -10.101010                         \\
-0.000010    & -101.100010   &  &  & -0.001000     & -11.100110     &  &  & -0.100000     & -1.110001                          \\
-0.000011    & -100.111110   &  &  & -0.001100     & -11.000010     &  &  & -0.110000     & -1.010011                          \\ \cline{1-2} \cline{5-6}  \cline{9-10} 
\end{tabular}
\captionof{table}{Illustration of co-transformation via an example. LNS design is represented by fixed-point numbers with 6 fractional bits.  
The three tables $T_a, T_b, T_c$ in binary with $\Delta_a = 0.000001$, $\Delta_b = 0.0001$, $\Delta_c = 0.01$.}
\label{tab:cotrans-illustration}
\end{center}
Suppose that we want to compute $\Phi^-(x)$ with $x=-0.000101$.
Because $x \in [-\Delta_b, -\Delta_a)$, we apply Case 2's formulae for calculation:
\begin{compactenum}

\item
Firstly, we compute $r_b$ and $r_a$ from the value of $x$ and $\Delta_b$, and get:
$r_b = -0.001000 $ and $r_a = -0.000011$. 
\item
Before computing $k$, we need to look up the values of $\Phi^-(r_a)$ and $\Phi^-(r_b)$
from Table~\ref{tab:cotrans-illustration}:
$\Phi^-(r_a) = -100.111110$ and $\Phi^-(r_b) = -11.100110$.
Then, we compute $k = x -\Phi^-(r_b)+\Phi^-(r_a) = -1.011101 $.
\item 
Next, we compute $\Phi^-(k)$ using first-order Taylor approximation (with $\Delta = 0.01$): 
$\hat{\Phi}_T^-(k) = -0.101001$
\item 
Finally, we compute  $\Phi^-(x)$ using Equation~\ref{eq:CT2}: 
$$\Phi^-(x) \approx \Phi^-(r_b) + \hat{\Phi}_T^-(k)= (-11.100110) + (-0.101001)  = -100.001111$$ 
\end{compactenum}
\end{cotransexample}

\paragraph{Error Analysis Conventions}
In this paper, we focus on deriving the 
error-bounds of approximating $\Phi^+$ and $\Phi^-$ using the 
first-order Taylor approximation, the error-correction technique
and the co-transformation techniques.
The error-bounds are parameterized by that machine-epsilon, $\epsilon$,
together with the parameters, which involves
in the calculation of the three techniques, such as $\Delta,\Delta_p, \Delta_a$ and $\Delta_b$.
There are many other designs of LNS, such as non-$2^\frac{1}{k}$
based~\cite{LNSbeyond2,madam},
varied $\Delta$~\cite{EM}, ROM-less~\cite{mitchellLNS,improveMitchell,ROMlessLNS,polyLNS}.
Covering all those designs is beyond the scope of this paper.
However, in Section~\ref{sec:conc}, we will show how the error-bound can be slightly 
modified for some popular LNS designs.

\section{High Level Overview of Entire Error Analysis}
\label{sec:high-level-overview-math}
In this section, we consecutively analyze all the sources that cause error
of computing $\Phi^+$ and $\Phi^-$ by the three techniques mentioned
in Section~\ref{sec:background}: first-order Taylor approximation,
error-correction, and co-transformation.
Then, we provide an overview of 
how the derivation of the 
rigorous error-bound for each technique proceeds, 
which will be described in detail 
in Sections~\ref{sec:Taylor}, ~\ref{sec:EC},  ~\ref{sec:cotrans},
giving each error mnemonic names (\S\ref{ss:FT})
\footnotemark
\footnotetext{We exclude the range $(-1,0)$ for $\Phi^-(x)$,
as this function tends to
negative infinity, and
hence
co-transformation 
is needed to meaningfully handle this range; this will be detailed in Section~\ref{sec:cotrans}).}

\subsection{Error analysis for First-Order Taylor Approximation}
\label{ss:FT}

Earlier, in Equation~\ref{eq:T}, we 
discussed how to mathematically calculate $\Phi^+$ and $\Phi^-$ 
via first-order Taylor approximation, 
the error of which was illustrated in Figure~\ref{fig:errortaylor}.
However, because LNS is implemented in hardware using fixed-point numbers, 
there are two more sources of error due to the fact that
the look-up tables' values of $\Phi(\mathbf{i})$ and $\Phi'(\mathbf{i})$
must be rounded to the current LNS's fixed-point representation, 
and the multiplication $\mathbf{r} \Phi'(\mathbf{i})$ is performed in
fixed-point arithmetic (with rounding).
Taking into account the implementation using fixed-point,
we refine the first-order Taylor approximation presented in Equation~\ref{eq:T} as:
\begin{equation}\label{eq:Ti}
\tilde{\Phi}_T(x) = \overline{\Phi}(\mathbf{i}) - \text{rnd}(\mathbf{r}\overline{\Phi'}(\mathbf{i}))     
\end{equation}
where $\overline{\Phi}$ and $\overline{\Phi'}$ are the fixed-point rounded 
look-up tables for $\Phi$ and $\Phi'$.\\
We define the notations for the three sources of error:
%
\begin{itemize}
\item Interp-err: the mathematical error of interpolating $\Phi^+$ and $\Phi^-$ 
via first-order Taylor approximation,
which is $|\Phi(x) - \hat{\Phi}_T(x)|$

\item Tab-err: the rounding error of the pre-computed values in the look-up tables,
which are $|\Phi(\mathbf{i}) - \overline{\Phi}(\mathbf{i})|$ and 
 $|\Phi'(\mathbf{i}) - \overline{\Phi'}(\mathbf{i})|$.

\item Mul-err: the error of fixed-point arithmetic multiplication,
which is $|\mathbf{r}\overline{\Phi'}(\mathbf{i}) - \text{rnd}(\mathbf{r}\overline{\Phi'}(\mathbf{i}))|$.

\end{itemize}
The error of interpolating $\Phi$ using first-order Taylor approximation is $|\Phi(x) - \tilde{\Phi}_T(x)|$,
where $\tilde{\Phi}_T(x)$ is defined in Equation~\ref{eq:Ti}.
The rigorous error-bound is derived as follows.

\paragraph{Error-bound of Interp-err derivation:}

Lemma~\ref{lem01} and Lemma~\ref{lem02} 
consecutively derive the error-bound of the Interp-err for $\Phi^+$ and $\Phi^-$, 
without considering Tab-err and Mul-err.
Intuitively, from the shape of Interp-err of $\Phi^+$ 
(illustrated in Figure~\ref{fig:errortaylor}), we observe that: 
\begin{compactitem}
    \item For each $\Delta$-segment, the error is greater when $x$ is further away from $0$.
    \item The further a $\Delta$-segment is away from $0$, the smaller its supremum error is.
\end{compactitem}
Lemma~~\ref{lem01} formally proves those observations by analyzing the error-function 
$|\Phi^+ (x)  - \hat{\Phi^+}_T(x) |$ and conclude that the supremum of the error over
the whole domain $x<0$ is approached when $x\to -\Delta$. 
Similarly, Lemma~~\ref{lem02} proves that the error-bound of $\Phi^-$ 
is obtained when $x\to -1-\Delta$. 
\paragraph{Total error-bound derivation:}

The error-bound of Tab-err and Mul-err is simply a constant $\epsilon$, 
which is the maximum absolute rounding error of the LNS' fixed-point representation.
The total error-bound is derived in Theorem~\ref{theo1} by accumulating (using absolute-norm inequality)
the error-bound of Interp-err with that of Tab-err and Mul-err.
%




\subsection{Error analysis for the Error Correction technique}
\label{ss:EC}

\begin{figure}[H]
\includegraphics[scale=1]{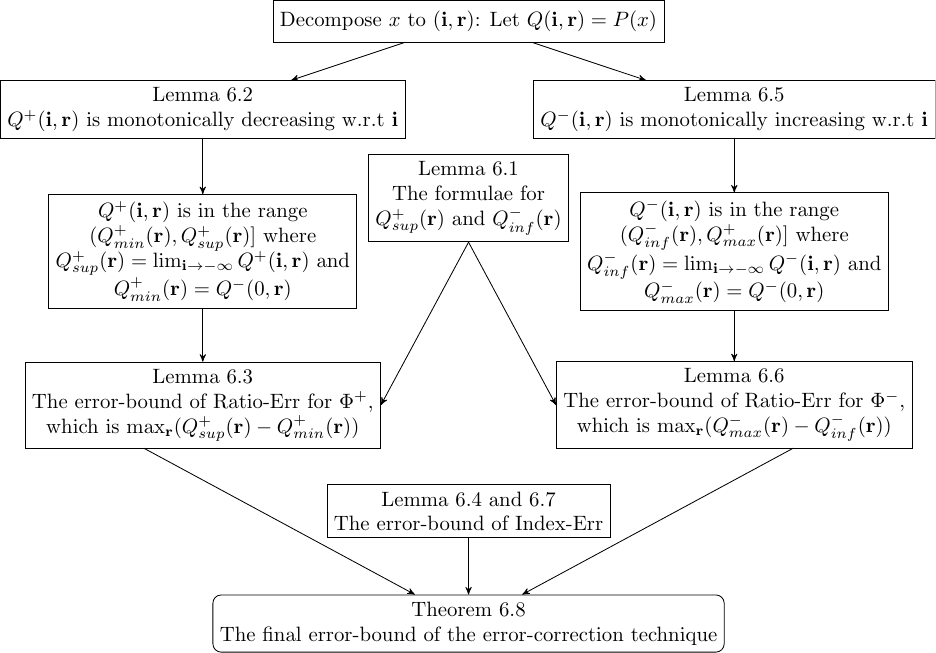}
  \caption{Derivation diagram of the rigorous error-bound of error correction technique}
  \label{fig:ecdiagram}
\end{figure}

Similar to the previous section, 
before deriving the error-bound,
we refine the mathematical formula of interpolation via error-correction technique (Equation~\ref{eq:EC})
such that it matches with the hardware implementation.
In hardware implementation, besides the fact that the look-up tables' values 
and results of multiplications are rounded 
according to the LNS' fixed-point representation,
we also note that 
the index values of $P_c$ are evenly spaced with distance $\Delta_P$
and any fixed-point number $\mathbf{r}$ is rounded-down to one of these indices $\hat{\mathbf{r}}$ before indexing.
Therefore, the formula
of interpolation via error-correction technique (Equation~\ref{eq:EC}) is {\em refined} as:
\begin{equation}\label{eq:ECi}
\tilde{\Phi}_{EC}(x) = \overline{\Phi}(\mathbf{i}) +\text{rnd}(\mathbf{r}\overline{\Phi'}(\mathbf{i})) - 
\text{rnd}(\overline{E_\Delta}(\mathbf{i})\overline{P_c}(\hat{\mathbf{r}}))    
\end{equation}
where $\overline{\Phi}$, $\overline{\Phi'}$, $\overline{P_c}$,
 $\overline{E_\Delta}$ are the fixed-point rounded 
look-up tables for $\Phi$, $\Phi'$,$P_c$ and $E_\Delta$.
\\
Next, we analyze the error sources of the refined formula of interpolation via the error-correction technique.
The main error of interpolation via the error-correction technique is that of
approximating the exact value of the ratio $P(x)$ by the pre-computed term $P_c(\mathbf{r})$
(mentioned earlier in Section~\ref{sec:background}).
Together with the hardware implementation errors above, we define the four sources of error (by splitting Inter-error into two parts shown below):
\begin{compactitem}
\item Interp-error decomposed into:

\begin{itemize}
\item Ratio-err: the error of approximating $P(x)$ by $P_c(\mathbf{r})$, which is $|P(x) - P_c(\mathbf{r})|$.
\item Index-err: the error caused by rounding $\mathbf{r}$ to an index value $\hat{\mathbf{r}}$ of the table $P_c$ before indexing, which is $|P_c(\mathbf{r}) - P_c(\hat{\mathbf{r}})|$
\end{itemize}

\item Tab-err: the rounding error of the pre-computed values in the look-up tables.
\item Mul-err: the rounding error of fixed-point arithmetic multiplication.
\end{compactitem}
The error of interpolating $\Phi$ using the error-correction technique is $|\Phi(x) - \tilde{\Phi}_{EC}(x)|$,
where $\tilde{\Phi}_{EC}(x)$ is defined in Equation~\ref{eq:ECi}.
The derivation of the rigorous error-bound of interpolation
using error-correction technique is fully described in Section~\ref{sec:EC}.
The structure its proof and the roles of the supporting lemmas are shown in Figure~\ref{fig:ecdiagram}.
The main idea of the proof is summarized as follows.\\

\begin{figure}[H]
\includegraphics[scale=1 ]{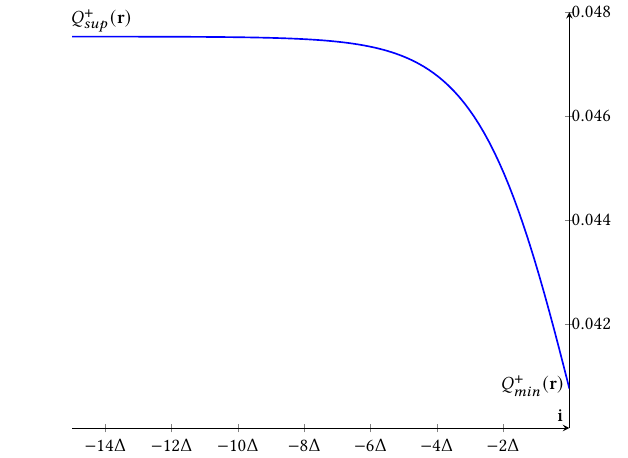}
  \caption{The range of $Q^+(\mathbf{i},\mathbf{r})$ w.r.t $\mathbf{i}$ where $\mathbf{r} = 0.2\Delta$}
  %
 \label{fig:additionratio}
\end{figure}

\paragraph{Error-bound of Ratio-err derivation:}
We observe that
deriving the
rigorous error
bound for 
Ratio-err
is fairly involved, and
proceeds as follows.
To facilitate
our analysis,
we define $Q(\mathbf{i},\mathbf{r}) = P(x)$ 
(see Section~\ref{sec:background} for the definition of $\mathbf{i}$ and $\mathbf{r}$).
Intuitively, for $\Phi^+$, we observe that for any value of $\mathbf{i}$, 
the value of $Q^+(\mathbf{i},\mathbf{r})$ is inside a finite range 
$[ Q^+_{min}(\mathbf{r}), Q^+_{sup}(\mathbf{r}))$,
where $Q^+_{sup}(\mathbf{r}) = \lim_{\mathbf{i} \to -\infty}Q^+(\mathbf{i},\mathbf{r})$ 
and $Q^+_{min}(\mathbf{r}) = Q^+(0,\mathbf{r})$ (illustrated in Figure~\ref{fig:additionratio}).
This observation is formally proved by 
Lemma~\ref{lem1}, which derives the formula of $Q^+_{sup}(\mathbf{r})$,
and Lemma~\ref{lem2}, which proves that $Q^+(\mathbf{i},\mathbf{r})$ is monotonically 
decreasing with respect to $\mathbf{i}$.

\begin{figure}[H]
\includegraphics[scale=1]{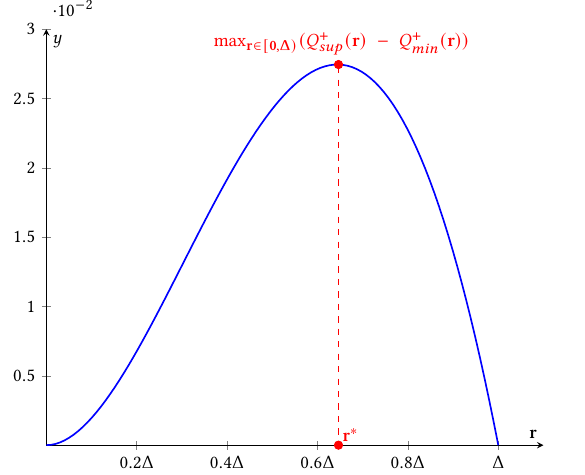}
  \caption{$Q^+_{sup}(\mathbf{r}) - Q^+_{min}(\mathbf{r})$}
  \label{fig:Fr}
\end{figure}

Then, obviously, the error-bound of the Ratio-err is the maximum value of 
$Q^+_{sup}(\mathbf{r}) - Q^+_{min}(\mathbf{r})$  
with respect to $\mathbf{r}$.
The graph of $Q^+_{sup}(\mathbf{r}) - Q^+_{min}(\mathbf{r})$ is 
illustrated in Figure~\ref{fig:Fr}.
Lemma~\ref{lem3} proves that the function
has a single maximum value at  $\mathbf{r^*}$ 
in the range $[0,\Delta]$ and
derives the formula for $\mathbf{r^*}$.
The error-bound of of the Ratio-err is 
obtained by substituting $\mathbf{r}$ with $\mathbf{r^*}$ in $Q^+_{sup}(\mathbf{r}) - Q^+_{min}(\mathbf{r})$.\\

\paragraph{Error-bound of Index-err derivation:}
Lemma~\ref{lem4} derives the error-bound of Index-err.
The key idea of the proof is to show that 
the supremum of $|Q^+(c,\mathbf{r})-P^+_c(\hat{\mathbf{r}})|$ is approached 
when $\mathbf{r}\to \Delta$ is rounded to $\hat{\mathbf{r}}=\Delta - \Delta_P$.
This idea comes from the observation that the error-ratio curve (see Figure~\ref{fig:errorratio}) 
of each $\Delta$-segment 
is steepest at the furthest point away from $0$
(mathematically, the second derivative of the $Q^+(c,\mathbf{r})$ w.r.t $\mathbf{r}$ is negative).
Similar steps are done for $\Phi^-$ by Lemma~\ref{lem5}, Lemma~\ref{lem6}, and Lemma~\ref{lem7}.

\paragraph{Total error-bound derivation:}
Finally, Theorem~\ref{the2} derives the formula for the total error-bound of $\tilde{\Phi}_{EC}(x)$, 
which considers all four sources of errors mentioned above. 
The total error-bound is derived by accumulating 
the error-bounds of Ratio-err and Index-err (derived in Lemma~\ref{lem3}, Lemma~\ref{lem4} for $\Phi^+$ 
and in Lemma~\ref{lem6}, Lemma~\ref{lem7} for $\Phi^-$) 
with the error-bounds of Tab-err and Mul-err (these two errors are simply the machine-epsilon $\epsilon$).

%

%

\subsection{Error analysis for the Co-transformation technique}
\label{ss:CT}
The calculation of all 3 cases of co-transformation technique 
mentioned at the end of Section~\ref{sec:background} involves 5 types of operations/computations:
div, subtraction, multiplication (of an integer and a fixed-point number),
indexing the value of $\Phi^-$ by table look-up, 
and computing $\Phi^-$ by interpolation 
(using first-order Taylor approximation or error-correction technique).
The first 3 operations are error-free in fixed-point arithmetic.
Indexing the value of $\Phi^-$ by table look-up has the error-bound $\epsilon$.
The error-bounds of computing $\Phi^-$ by interpolation is derived 
in subsections~\ref{ss:FT} and~\ref{ss:EC}
(details in Section~\ref{sec:Taylor} and~\ref{sec:EC}),
but with the assumption that the input of the $\Phi^-$ function is error-free.
However, in the co-transformation technique's calculation, 
 $\Phi^-(k)$ (Case 2), $\Phi^-(k_1)$ (Case 3) and $\Phi^-(k_2)$ (Case 3) are computed
by interpolation while the computations of $k, k_1$ and $k_2$ contain errors.
To address this challenge, we derive the error-bound of $\Phi^-(k)$ from the error-bound of $k$ 
assuming that the computation of $\Phi^-$ is exact (Lemma ~\ref{lem31}),
then combine it with the derived error-bound of computing $\Phi^-$ by one of the two interpolation
techniques.
%
Finally, Theorem~\ref{theorem3} derives to total error-bound of computing 
$\Phi^-$ in the range $(-1,0)$ using the co-transformation technique from
the error-bounds of the three cases:
\begin{compactenum}
\item For Case 1, there is only one look-up table indexing, so the error-bound is simply $\epsilon$ 
\item For Case 2, $r_b$ and $r_a$ are error-free,
the error-bound of computing $k$ is $2\epsilon$ 
(two look-up table indexing of $\Phi^-(r_a)$ and $\Phi^-(r_b)$).
Next, the error-bound of computing $\Phi^-(k)$ is derived directly by applying Lemma ~\ref{lem31}.
Finally, the total error-bound of computing $\Phi^-(x)$ in Case 2 
follows by accumulating one more $\epsilon$, 
which is the error-bound of indexing $\Phi^-(r_b)$.
\item For Case 3, deriving the error-bounds of computing $k_1$ and $\Phi^-(k_1)$ is similar to 
that of $k$ and $\Phi^-(k)$ in Case 2. 
The error-bound of computing $k_2$ is obtained by accumulating error-bound of computing $\Phi^-(k_1)$
with $2\epsilon$ (two look-up table indexing of $\Phi^-(r_a)$ and $\Phi^-(r_b)$).
Similar to Case 2, the error-bound of computing $\Phi^-(k_1)$ is derived by applying Lemma ~\ref{lem31},
and the total error-bound  computing $\Phi^-(x)$ in Case 3 
follows by accumulating one more $\epsilon$, 
which is the error-bound of indexing $\Phi^-(r_c)$.
\end{compactenum}
The error-bound of Case 1 is obviously smaller than those of Case 2 and Case 3,
so the total error-bound of computing 
$\Phi^-$ in the range $(-1,0)$ using co-transformation technique 
is the maximum of the error-bound of Case 2 and Case 3.

\section{Rigorous error bound of first-Order Taylor approximation}

\label{sec:Taylor}
This section mentions in full detail the derivation of the error-bound of calculating $\Phi^+$ and $\Phi^-$
using first-order Taylor approximation. 
The structure of this section and the intuition of the proof were briefly mentioned in Section~\ref{sec:high-level-overview-math}.1:

Lemma~\ref{lem01} derive the error-bound of Interp-err of $\Phi^+$ in the range $(-\infty,0]$.
Recall that the Interp-err is $|\Phi^+ (x)  - \hat{\Phi^+}_T(x) | $ (plotted in
Figure~\ref{fig:errortaylor}) and 
$E_\Delta(\mathbf{i}) = \Phi(\mathbf{i}-\Delta) -\Phi (\mathbf{i})  + \Delta\Phi'(\mathbf{i})$
(defined in Section~\ref{sec:background}).
\begin{lemma}\label{lem01}
For all $ x\in (-\infty,0]$,
$$|\Phi^+ (x)  - \hat{\Phi^+}_T(x) | 
 \leq  E^+_\Delta(0) $$
\end{lemma}

\begin{proof}
Let 
$E(\mathbf{i},\mathbf{r}) = \Phi^+ (x)  - \hat{\Phi^+}_T(x)$.
From the definition of $\hat{\Phi^+}_T(x)$ in Equation~\ref{eq:T}:
$$E(\mathbf{i},\mathbf{r}) =  \Phi^+(\mathbf{i}-\mathbf{r}) - (\Phi^+(\mathbf{i}) - \mathbf{r}(\Phi^+)'(\mathbf{i}))$$
Note that despite the definition: $\mathbf{i}=\Delta(x \text{ div }  \Delta)$,
it is safe to consider the domain of $\mathbf{i}$ to be $\mathbb{R}_{\leq 0}$ 
instead of $\Delta\mathbb{Z}_{\leq0}$  when deriving error-bound,
%
%
because:
$$\max_{\mathbf{i}\in \Delta \mathbb{Z}_{\leq0}, 0 \leq \mathbf{r} <\Delta} |E(\mathbf{i},\mathbf{r})|
\leq \max_{\mathbf{i} \in \mathbb{R}_{\leq 0}, 0 \leq \mathbf{r} <\Delta} |E(\mathbf{i},\mathbf{r})| 
\quad \quad (\text{because } \Delta\mathbb{Z}_{\leq0} \subset \mathbb{R}_{\leq 0})$$
\\
The lemma is proved by:
\begin{compactenum}
    \item 
    Proving that $\forall \mathbf{i} \leq 0, 0 \leq \mathbf{r} < \Delta: E(\mathbf{i},\mathbf{r}) \geq 0 $
    , so $E(\mathbf{i},\mathbf{r})  = |E(\mathbf{i},\mathbf{r})| = |\Phi^+ (x)  - \hat{\Phi^+}_T(x) | $

    \item Proving that both partial derivatives of $E$ w.r.t $\mathbf{r}$ and $\mathbf{i}$
    are non-negative, so $E$ approaches its supremum when $\mathbf{i}=0$ and $\mathbf{r}\to \Delta$.
    Formally, we have to
    prove that $\forall \mathbf{i} \leq 0, 0 \leq \mathbf{r} < \Delta:
    \frac{\partial E}{\partial \mathbf{r}}(\mathbf{i},0)
    \geq 0$ and $\frac{\partial E }{\partial \mathbf{i}}(\mathbf{i},0) \geq 0$, 
    so $\max_{\mathbf{i} \in \mathbb{R}_{\leq 0}, 0 \leq \mathbf{r} <\Delta} E(\mathbf{i},\mathbf{r}) <
    E(0,\Delta) = E^+_\Delta(0) $.
\end{compactenum}
First, we take the first and second derivatives of $E(\mathbf{i},\mathbf{r})$ w.r.t  $\mathbf{r}$:
$$ \frac{\partial E }{\partial \mathbf{r}}(\mathbf{i},\mathbf{\mathbf{r}}) = 
\frac{2^\mathbf{i}}{2^\mathbf{i}+1} - \frac{2^{\mathbf{i}-\mathbf{r}}}{2^{\mathbf{i}-\mathbf{r}}+1}   
\quad \text{and} \quad
\frac{\partial^2 E }{\partial \mathbf{r}^2}(\mathbf{i},\mathbf{\mathbf{r}}) = 
\frac{ 2^{\mathbf{i}-\mathbf{r}} \ln 2 }{(2^{\mathbf{i}-\mathbf{r}}+1)^2} $$
From $\frac{\partial^2 E }{\partial \mathbf{r}^2} (\mathbf{i},\mathbf{r}) > 0$ and $\frac{\partial E }{\partial \mathbf{r}}(\mathbf{i},0)=0$, we conclude that $ \forall \mathbf{i} \leq 0, 0 \leq \mathbf{r} < \Delta:
 \frac{\partial E }{\partial \mathbf{r}}(\mathbf{i},\mathbf{r}) \geq 0$.
\\
Then, because $E (\mathbf{i},0) =0$, we conclude that $ \forall \mathbf{i} \leq 0, 0 \leq \mathbf{r} < \Delta:
 E (\mathbf{i},\mathbf{\mathbf{r}}) \geq 0$. 
 \footnote{Since  $\frac{\partial^2 E }{\partial \mathbf{r}^2} (\mathbf{i},\mathbf{r}) 0$
 is strictly greater than $0$, we conclude that if $\mathbf{r}>0$ then $E (\mathbf{i},\mathbf{\mathbf{r}}) > 0$
 \label{fn:Egt0}}\\
We complete the proof by proving that $\forall \mathbf{i} \leq 0, 0 \leq \mathbf{r} < \Delta: 
\frac{\partial E}{\partial \mathbf{i}} (\mathbf{i},\mathbf{\mathbf{r}}) \geq 0:$
\footnote{When $\mathbf{r}>0$, we can replace all $\geq$ and $\leq$ in the rest of the proof by $>$ and $<$ correspondingly
to prove that $\frac{\partial E}{\partial \mathbf{i}} (\mathbf{i},\mathbf{\mathbf{r}}) > 0$
\label{fn:Eigt0}}
\\
If $\mathbf{i} \leq 0, 0 \leq \mathbf{r} < \Delta$, let $a= \mathbf{r}\ln 2 $, then
$$ \frac{\partial E}{\partial \mathbf{i}} (\mathbf{i},\mathbf{\mathbf{r}}) = 
\frac{2^\mathbf{i}}{(2^\mathbf{i}+1)^2 (2^{\mathbf{i}-\mathbf{r}}+1)}
(2^\mathbf{i} f(a) + g(a))$$
$$\text{with} \quad f(a) = a e^{-a} + e^{-a} - 1 \quad \text{and} \quad
 g(a) = e^{-a} + a - 1
$$
\\
Since $\frac{2^\mathbf{i}}{(2^\mathbf{i}+1)^2 (2^{\mathbf{i}-\mathbf{r}}+1)}>0$,
the sign of $ \frac{\partial E}{\partial \mathbf{i}} (\mathbf{i},\mathbf{\mathbf{r}})$ is the same as that of
$ N(\mathbf{i}) = 2^\mathbf{i} f(a) + g(a)$
\\
Because $a \geq 0$, from  $f(0)=0$ and $f'(a) = -ae^{-a} \leq 0 $,
we conclude that $f(a) \leq 0 $, so
$$N'(\mathbf{i}) = 2^\mathbf{i} (\ln 2) f(a) \leq 0 $$
Let $h(a) = N(0) = (a+2)e^{-a} + a-2$, then 
$h(0) = 0$ and $h'(a) = -f(a) \geq 0$,
we conclude that: $h(a) = N(0) \geq 0$.\\
From $N(0) \geq 0$ and $N'(\mathbf{i}) \leq 0$, we conclude that
for all $\mathbf{i} \leq 0$, $N(\mathbf{i}) \geq N(0) \geq 0$. \\
Hence, $\forall \mathbf{i} \leq 0, 0 \leq \mathbf{r} < \Delta: \frac{\partial E}{\partial \mathbf{i}} (\mathbf{i},\mathbf{\mathbf{r}}) \geq 0$

\end{proof}

Lemma~\ref{lem02} derives the error-bound of Interp-err of $\Phi^-$ in the range $(-\infty,-1]$. 
Note that for the range $(-1,0)$, $\Phi^-$ is calculated by the co-transformation technique,
the error-bound of which is derived in Section~\ref{sec:cotrans}.
\begin{lemma}\label{lem02}
$\forall x\leq 0:$ 
$$|\Phi^- (x)  - \hat{\Phi^-}_T(x) | 
 \leq - E^-_\Delta(-1) $$
\end{lemma}

\begin{proof}
The proof is very similar to that of Lemma~\ref{lem01}
See Appendix~\ref{ss:prooflem52} for proof's details.
\end{proof}

Theorem~\ref{theo1} derives the total error-bound of computing $\Phi^+$ and $\Phi^-$
using first-order Taylor approximation.
\begin{theorembox}
\begin{theorem} \label{theo1}
Let $\epsilon$ be the machine-epsilon of the fixed-point representation of
the LNS under consideration.
Let
 $E^+_M = E_\Delta^+(0)$,
 and $E^-_M = -E_\Delta^-(-1)$.
 Then
$$|\Phi(x) - \tilde{\Phi}_T(x) |  < E_M + (2+\Delta)\epsilon$$
\end{theorem}
\end{theorembox}
\begin{proof}
Applying the absolute-value norm inequality $|\sum a_i| \leq \sum |a_i|$:
$$|\Phi(x) - \tilde{\Phi}_T(x)  | \leq 
|\Phi(x) - \hat{\Phi}_T(x)  | + | \hat{\Phi}_T(x)  - \tilde{\Phi}_T(x) |  $$ 
$$= E_M +  | \hat{\Phi}_T(x)  - \tilde{\Phi}_T(x) |$$
From Equation~\ref{eq:T} and Equation~\ref{eq:Ti}, $\hat{\Phi}_T(x)  - \tilde{\Phi}_T(x)$
can be re-written as $a_1 + a_2 + a_3$, where
$$a_1 = \Phi(\mathbf{i}) - \overline{\Phi}(\mathbf{i})$$
$$a_2 = \mathbf{r}(\Phi'(\mathbf{i}) - \overline{\Phi'}(\mathbf{i}))$$
$$a_3 = \mathbf{r}\overline{\Phi'}(\mathbf{i}) - \text{rnd}(\mathbf{r}\overline{\Phi'}(\mathbf{i}))$$
Apply the absolute-value norm inequality $|\sum a_i| \leq \sum |a_i|$ again:
$$| \hat{\Phi}_T(x)  - \tilde{\Phi}_T(x) | \leq |a_1|+ |a_2| + |a_3|$$
From $|a_1|, |a_3|, |\Phi'(\mathbf{i}) - \overline{\Phi'}(\mathbf{i})| \leq \epsilon$
(as they are errors of fixed-point rounding) and $0\leq \mathbf{r} < \Delta $:
$$| \hat{\Phi}_T(x)  - \tilde{\Phi}_T(x) | < (2+\Delta) \epsilon$$
Hence, the final error-bound of first-order Taylor approximation is obtained as:
$$|\Phi(x) - \tilde{\Phi}_T(x) |  < E_M + (2+\Delta)\epsilon$$
\end{proof}

\section{Rigorous error bound of the error correction technique}
\label{sec:EC}
This section mentions in full detail the derivation of the rigorous error-bound of computing $\Phi^+$ and $\Phi^-$
using the error-correction technique, together with the formal mathematical proof. 
The structure of this section is shown in Figure~\ref{fig:ecdiagram}
and the intuition of the proof was briefly mentioned in Section~\ref{ss:EC}.
For a quick recall, there are four sources of error of interpolating by the error-correction technique:
Ratio-err, Index-err, Tab-err and Mul-err. 
The error-bound of Ratio-err are derived in Lemma~\ref{lem3} (for $\Phi^+$) and Lemma~\ref{lem6} (for $\Phi^-$).
The error-bound of Index-err are derived in Lemma~\ref{lem4} (for $\Phi^+$) and Lemma~\ref{lem7} (for $\Phi^-$).
Finally, Theorem~\ref{the2} accumulates the error-bounds of Ratio-err and Index-err 
with those of Tab-err and Mul-err to derives the total error-bound.

Recall that we defined $Q(\mathbf{i},\mathbf{r}) = P(\mathbf{i} - \mathbf{r}) = P(x)$.
Similar to the $E(\mathbf{i},\mathbf{r})$ in the proof of Lemma~\ref{lem01},
for the sake of deriving the error-bound, it is safe to consider the domain of $\mathbf{i}$
to be $\mathbb{R}_{\leq 0}$. 
Lemma~\ref{lem1} derives the limit of $Q^+(\mathbf{i},\mathbf{r})$ and $Q^-(\mathbf{i},\mathbf{r})$
when $\mathbf{i}\to -\infty$, which later will be proved to be the supremum of $Q^+(\mathbf{i},\mathbf{r})$ 
and infimum of $Q^-(\mathbf{i},\mathbf{r})$. 

\begin{lemma}\label{lem1}
$$\lim_{\mathbf{i} \to -\infty} Q^+(\mathbf{i},\mathbf{r}) 
= \lim_{\mathbf{i} \to -\infty} Q^-(\mathbf{i},\mathbf{r})
= \frac{2^{-\mathbf{r}} + \mathbf{r}\ln 2 - 1}{2^{-\Delta} + \Delta\ln 2 -1}$$
\end{lemma}
\begin{proof}
From the formula of $P(x)$:
\begin{equation}\label{eq:QE}
Q^+(\mathbf{i},\mathbf{r}) = P^+ (\mathbf{i} - \mathbf{r}) =  \frac{{E^+}_T(\mathbf{i} - \mathbf{r})}{{E^+}_\Delta(\mathbf{i})}
= \frac{\Phi^+ (\mathbf{i} - \mathbf{r}) - \Phi^+(\mathbf{i}) + \mathbf{r} (\Phi^+)'(\mathbf{i})}
{\Phi^+ (\mathbf{i}-\Delta) - \Phi^+(\mathbf{i}) + \Delta (\Phi^+)'(\mathbf{i})}.
\end{equation}
Because $(\Phi^+)'(\mathbf{i}) = \frac{2^{\mathbf{i}}}{2^{\mathbf{i}}+1}$, 
we rewrite $Q^+(\mathbf{i},\mathbf{r})$ as:
\begin{equation}\label{eq:Q}
Q^+(\mathbf{i},\mathbf{r}) =\frac{2^\mathbf{i} \mathbf{r} \ln 2 - (2^\mathbf{i} +1) \ln (2^\mathbf{i}+1) + (2^\mathbf{i} +1) \ln (2^{\mathbf{i}-\mathbf{r}}+1)}
{2^\mathbf{i} \Delta \ln 2 - (2^\mathbf{i} +1) \ln (2^\mathbf{i}+1) + (2^\mathbf{i} +1) \ln (2^{\mathbf{i}-\Delta}+1)}.    
\end{equation}
Define $f(a, r) = a r \ln 2 - (a + 1) \ln (a+1) + (a +1) \ln (a 2^{-r}+1)$. Then
$$
Q^+(\mathbf{i}, \mathbf{r}) 
  = Q_a^+(a, \mathbf{r})
  = \frac{f(2^{\mathbf{i}}, \mathbf{r})}{f(2^{\mathbf{i}}, \Delta)}.
$$
We have $\lim_{a \to 0} f(a, r) = f(0, r) = 0$ and
$$
\lim_{a \to 0} \frac{\partial f(a, r)}{\partial a} 
  = \lim_{a \to 0} \left(r \ln 2 - \ln (a + 1) - 1 + \ln (a 2^{-r} + 1) 
        + \frac{2^{-r}(a + 1)}{a 2^{-r} + 1}\right)
  = r \ln 2 - 1 + 2^{-r}.
$$
Since $\lim_{\mathbf{i} \to -\infty} 2^{\mathbf{i}} = 0$ and $2^{\mathbf{i}} > 0$ for all $\mathbf{i}$, we can compute the limit of $Q^+(\mathbf{i},\mathbf{r})$ with L'Hopital's rule as follows
$$
\lim_{\mathbf{i} \to -\infty} Q^+(\mathbf{i},\mathbf{r}) 
  = \lim_{a \to 0^+} \frac{f(a, \mathbf{r})}{f(a, \Delta)}
  = \lim_{a \to 0^+} \frac{\frac{\partial}{\partial a}f(a, \mathbf{r})}
                          {\frac{\partial}{\partial a}f(a, \Delta)}
  = \frac{2^{-\mathbf{r}} + \mathbf{r}\ln 2 - 1}{2^{-\Delta} + \Delta\ln 2 - 1}.
$$
Similarly, we get the same result for $\lim_{\mathbf{i}\to -\infty}Q^-(\mathbf{i},\mathbf{r})$
\end{proof}
Lemma~\ref{lem2} proves that $Q^+(\mathbf{i},\mathbf{r})$ is monotonically decreasing w.r.t $\mathbf{i}$
(its partial derivative over $\mathbf{i}$ is negative), when $\mathbf{i} \leq 0$. 
The purpose of this Lemma is to show that the value of $Q^+(\mathbf{i},\mathbf{r})$ is inside the range
$[Q_{min}^+(\mathbf{r}), Q^+_{sup}(\mathbf{r}) )$ where $Q_{min}^+(\mathbf{r})=Q^+ (0,\mathbf{r})$
and $Q^+_{sup}(\mathbf{r}) = \lim_{\mathbf{i} \to -\infty} Q^+(\mathbf{i},\mathbf{r}) $.
The range is shown in Figure~\ref{fig:additionratio}; 
the small numerical value of the range verifies the observation that the error-ratio function (Figure~\ref{fig:errorratio})
is roughly periodic.
\begin{lemma} \label{lem2}
 $\forall \mathbf{i}\leq 0$, $0\leq \mathbf{r}<\Delta: \quad Q^+(\mathbf{i},\mathbf{r}) \text{ is monotonically decreasing w.r.t  } \mathbf{i}$

\end{lemma}

\begin{proof}
We prove the lemma by proving that  
$\forall \mathbf{i} < 0$, $0 < \mathbf{r}<\Delta$: $Q^+(\mathbf{i},\mathbf{r})$ is continuous and
$\frac{\partial Q^+(\mathbf{i},\mathbf{r})}{\partial \mathbf{i}} \leq 0$.

Similar to Lemma~\ref{lem01}, 
we define  $E(\mathbf{i},\mathbf{r}) = \Phi^+(\mathbf{i}-\mathbf{r}) - (\Phi^+(\mathbf{i}) - \mathbf{r}(\Phi^+)'(\mathbf{i}))$.

From Equation~\ref{eq:QE}, proving that
$\forall \mathbf{i} < 0$, $0 < \mathbf{r}<\Delta$: $Q^+(\mathbf{i},\mathbf{r})$ is continuous
is equivalent to proving that $\forall \mathbf{i} < 0$: $E(\mathbf{i},\Delta) \neq 0$, which follows directly from
footnote~\ref{fn:Egt0} as $\Delta>0$.

For the rest of the proof, we prove that 
$\forall \mathbf{i} < 0$, $0 < \mathbf{r}<\Delta$:
$\frac{\partial Q^+(\mathbf{i},\mathbf{r})}{\partial \mathbf{i}} \leq 0$.

Let $h(\mathbf{i},\mathbf{r}) =  E(\mathbf{i},\mathbf{r}) (2^\mathbf{i} + 1 ) \ln 2 =
2^\mathbf{i} \mathbf{r} \ln 2 - (2^\mathbf{i} +1) \ln (2^\mathbf{i}+1) + (2^\mathbf{i} +1) \ln (2^{\mathbf{i}-\mathbf{r}}+1)$. 
Then from Equation~\ref{eq:Q}:
$$\frac{\partial Q^+(\mathbf{i},\mathbf{r})}{\partial \mathbf{i}} = 
\frac{\frac{\partial h}{\partial \mathbf{i}} (\mathbf{i},\mathbf{r}) h(\mathbf{i},\Delta) - \frac{\partial h}{\partial \mathbf{i}} (\mathbf{i},\Delta) h(\mathbf{i},\mathbf{r})}
{h(\mathbf{i},\Delta)^2}$$
Because
$$\frac{\partial Q^+(\mathbf{i},\mathbf{r})}{\partial \mathbf{i}} \leq  0
\Leftrightarrow \frac{h(\mathbf{i},\mathbf{r})}{\frac{\partial h}{\partial \mathbf{i}} (\mathbf{i},\mathbf{r})}
\geq \frac{h(\mathbf{i},\Delta)}{\frac{\partial h}{\partial \mathbf{i}} (\mathbf{i},\Delta)},$$
Let $g(\mathbf{i},\mathbf{r}) = \frac{h(\mathbf{i},\mathbf{r})}{\frac{\partial h}{\partial  \mathbf{i}} (\mathbf{i},\mathbf{r})}$. 
Since $\mathbf{r} < \Delta$,
we prove $\forall \mathbf{i} < 0$, $0 < \mathbf{r}<\Delta$:
$\frac{\partial Q^+(\mathbf{i},\mathbf{r})}{\partial \mathbf{i}} \leq 0$
by proving that  
$\forall \mathbf{i} < 0$, $0 < \mathbf{r}<\Delta$: 
$g(\mathbf{i},\mathbf{r})$ is continuous $^{(1)}$
and $\frac{\partial g}{\partial  \mathbf{r}} (\mathbf{i},\mathbf{r}) \leq 0$ $^{(2)}$.\\
Proving $^{(1)}$ is equivalent to proving that $\forall \mathbf{i} < 0$, $0 < \mathbf{r}<\Delta$:  $\frac{\partial h}{\partial  \mathbf{i}} (\mathbf{i},\mathbf{r}) \neq 0$  .

From $h(\mathbf{i},\mathbf{r}) =  E(\mathbf{i},\mathbf{r}) (2^\mathbf{i} + 1 ) \ln 2 $,
we derive $\frac{\partial h}{\partial  \mathbf{i}} (\mathbf{i},\mathbf{r}) $:
$$\frac{\partial h}{\partial  \mathbf{i}} (\mathbf{i},\mathbf{r}) 
= \left( \frac{\partial E}{\partial \mathbf{i}} (\mathbf{i},\mathbf{r}) (2^\mathbf{i} + 1 ) + 
 E(\mathbf{i},\mathbf{r}) 2^\mathbf{i} \ln 2
\right) 
\ln 2
$$
From the fact that if $\mathbf{r}>0$ then $ E(\mathbf{i},\mathbf{r}) >0 $ and $\frac{\partial E}{\partial \mathbf{i}} (\mathbf{i},\mathbf{r}) >0$
(see footnotes~\ref{fn:Egt0} and~\ref{fn:Eigt0} ), we conclude that $\forall \mathbf{i} < 0$, $0 < \mathbf{r}<\Delta$:  $\frac{\partial h}{\partial  \mathbf{i}} (\mathbf{i},\mathbf{r}) > 0$ and complete the proof for $^{(1)}$.\\
To prove $^{(2)}$, we derive $\frac{\partial g}{\partial  \mathbf{r}} (\mathbf{i},\mathbf{r})$: \footnotemark
\footnotetext{The derivative is calculated using Sympy library} 
$$\frac{\partial g}{\partial \mathbf{r}} (\mathbf{i},\mathbf{r})= \frac{(B-1)K}{A M^2} \quad$$
where
\begin{compactitem}
\item[•] $A = 2^\mathbf{i}$, $B = 2^\mathbf{r}$
\item[•] $C = \ln(A+1)$
\item[•] $D = \ln(A+B)$
\item[•] $M = AD-AC-BC+BD-B+1$
\item[•] $K = A^2\ln(A+B)  -A^2\ln (A+1) -AB + A  + B\ln B
+B\ln (A+1) - B\ln (A+B)$
\end{compactitem}
Because $A > 0$ and $B \geq 1$
 (since $ \mathbf{i} \leq 0, \mathbf{r}\geq 0$ ),
$\frac{\partial g}{\partial r} (\mathbf{i},\mathbf{r}) \leq 0$ is equivalent to $K \leq 0$ $^{(3)}$.
\\
Considering $K$ as a function of 2 variables $A$ and $B$,
we take the first and second derivatives of $K$ w.r.t $B$:
$$\frac{\partial K}{\partial B}(A,B)= \frac{A^2 -B}{A+B}
-A + \ln (A+1) + \ln B - \ln(A+B) +1$$
$$\frac{\partial^2 K}{\partial B^2}(A,B)= 
\frac{A^2(1-B)}{B(A+B)^2} $$
\\
Substituting $B$ for $1$ in $\frac{\partial K }{\partial B} $,
we get $\frac{\partial K }{\partial B}(A,1) =0$. 
Moreover, $\frac{\partial^2 K}{\partial B^2}(A,B) \leq 0$  since $A > 0$ and $B \geq 1$, 
we conclude that
 $\frac{\partial K }{\partial B}(A,B) \leq0$.
Similarly, substituting $B$ for $1$ in $K$, we get $K(A,1) =0$, 
together with $\frac{\partial K }{\partial B}(A,B) \leq0$, we conclude that $K(A,B)\leq0$ $^{(4)}$.
The result of this lemma follows directly from $^{(2)}$, $^{(3)}$, and $^{(4)}$.
 
\end{proof}
Lemma~\ref{lem3} derives the maximum of the range (i.e. $Q^+_{sup}(\mathbf{r}) - Q^+ (0,\mathbf{r})$) w.r.t $\mathbf{r}$, which is the error-bound of the Ratio-err.
\begin{lemma} \label{lem3}
$\forall \mathbf{i}\leq 0,$ $c \leq 0$, $\mathbf{r}\in [0, \Delta)$:
$$|Q^+(\mathbf{i},\mathbf{r}) - Q^+(c,\mathbf{r})| < Q^+_{sup}(\mathbf{r}^*) - Q^+_{min}(\mathbf{r}^*)$$
where
$$Q^+_{sup}(\mathbf{r}) = 
\lim_{i \to -\infty} Q^+(\mathbf{i},\mathbf{r})  =
\frac{2^{-\mathbf{r}} + \mathbf{r}\ln 2 - 1}{2^{-\Delta} + \Delta\ln 2 -1} $$
$$Q^+_{min}(\mathbf{r}) = Q^+(0,\mathbf{r})
=\frac{\mathbf{r}\ln 2 + 2 \ln (1+2^{-\mathbf{r}}) - 2 \ln 2}
{\Delta\ln 2 + 2 \ln (1+2^{-\Delta}) - 2 \ln 2} $$
$$\mathbf{r}^* = \log_2 \frac
{-X(2\ln (X+1) - \ln X - 2\ln 2) }
{ 2X(\ln (X+1) -\ln X - \ln 2) + X -1  } $$
where $X = 2^\Delta$
\end{lemma}

\begin{proof}
From Lemma~\ref{lem1} and Lemma~\ref{lem2}:
 $\forall \mathbf{i}\leq 0, \mathbf{r}\in [0, \Delta):$
$Q^+(\mathbf{i},\mathbf{r}) \in [Q^+_{min}(\mathbf{r}), Q^+_{sup}(\mathbf{r}))$,
then $\forall \mathbf{i}\leq 0,$ $c \leq 0$, $\mathbf{r}\in [0, \Delta)$:
$|Q^+(\mathbf{i},\mathbf{r}) - Q^+(c,\mathbf{r})|  < Q^+_{sup}(\mathbf{r}) - Q^+_{min}(\mathbf{r})$.\\
Let $F(\mathbf{r}) = Q^+_{sup}(\mathbf{r}) - Q^+_{min}(\mathbf{r})$,
proving the lemma is equivalent to proving that $\mathbf{r}^* = \text{argmax}_{[0,\Delta)} F $.
\\
We take the first derivative of $F$ \footnotemark:
\footnotetext{The derivative is calculated using Sympy library}
$$F'(\mathbf{r}) = \frac{X2^{-\mathbf{r}}(2^\mathbf{r} -1)\ln 2 }
{(2^\mathbf{r} + 1)(B-A)B}
\left(A2^\mathbf{r} + B \right)$$
where
$$A = 2X(\ln (X+1) -\ln X - \ln 2) + X -1 $$
$$B = X(2\ln (X+1) - \ln X - 2\ln 2) $$
Note that $\mathbf{r}^*=\log_2(-B/A)$.
Considering $A$ and $B$ as functions of $X$, since $X=2^\Delta >1 $,
performing basic function analysis, we can prove that
$A < 0$, $B>0$ and $B-A>0$.
Therefore, $F'(\mathbf{r}) = 0$ when $\mathbf{r}= \mathbf{r}^*$;
$F'(\mathbf{r}) > 0$ when $0<\mathbf{r}< \mathbf{r}^*$; and
$F'(\mathbf{r}) < 0$ when $\mathbf{r}> \mathbf{r}^*$ (1).
Because, $B-A>0$, $\mathbf{r}^*=\log_2(-B/A) >0$ (2).
Since $F(0) = F(\Delta)=0$, $\mathbf{r}^* < \Delta$; otherwise
$F(0) < F(\Delta)=0$, contradiction (3).
Hence, from (1), (2), and (3), we conclude that $\mathbf{r}^* = \text{argmax}_{[0,\Delta)} F$.
\end{proof}
Lemma~\ref{lem4} derives error-bound of the Index-err. 
\begin{lemma}\label{lem4}
Let  $\Delta_P$ be the distance of two adjacent index values of $P_c^+$
tables and $\hat{\mathbf{r}}$ be the rounding of $\mathbf{r}\in [0,\Delta)$ in modulo $\Delta_P$
, i.e., $\hat{\mathbf{r}} = \left\lceil \dfrac{x}{\Delta_P} \right\rceil \Delta_P$, then 
$$|Q^+(c,\mathbf{r})-P_c^+(\hat{\mathbf{r}})| \leq  1- Q_{min}^+(\Delta -\Delta_P)  $$
\end{lemma}
\begin{proof}
Let $E(\mathbf{i},\mathbf{r}) = \Phi(\mathbf{i}-\mathbf{r}) - (\Phi(\mathbf{i}) - \mathbf{r}\Phi'(\mathbf{i}))$,
then
$\frac{\partial Q^+}{\partial \mathbf{r}}(c,\mathbf{r}) = \frac{1}{E(c,\Delta)} \frac{\partial E}{\partial \mathbf{r}}(c,\mathbf{r}) \geq 0$ (see proof of Lemma~\ref{lem01}). 
Moreover because $\mathbf{r} > \hat{\mathbf{r}}$, we conclude that $ Q^+(c,\mathbf{r})-Q^+(c,\hat{\mathbf{r}})\geq 0$.
Therefore,
$|Q^+(c,\mathbf{r})-P_c^+(\hat{\mathbf{r}})| = 
Q^+(c,\mathbf{r})-P_c^+(\hat{\mathbf{r}})$,
because $ Q^+(c,\mathbf{r})-P_c^+(\hat{\mathbf{r}}) =  Q^+(c,\mathbf{r})-Q^+(c,\hat{\mathbf{r}}) \geq 0$.
\\
Let $t = \mathbf{r} -\hat{\mathbf{r}}$ and $F(\mathbf{r},t) = Q^+(c,\mathbf{r})-Q^+(c, \mathbf{r} -t)$, then
$$F(\mathbf{r},t) = |Q^+(c,\mathbf{r})-P_c^+(\hat{\mathbf{r}})|=
\frac{E(c,\mathbf{r})-E(c,\mathbf{r}-t)}{E(c,\Delta)} $$
Since $0\leq \mathbf{r}<\Delta$ and $0\leq  t =\mathbf{r}-\hat{\mathbf{r}} <\Delta_P$:
$$F(\mathbf{r},t) = |Q^+(c,\mathbf{r})-P_c^+(\hat{\mathbf{r}})|
< \max_{0\leq t\leq \Delta_P, t\leq \mathbf{r}\leq \Delta} F(\mathbf{r},t) $$
We will prove that $\max_{0\leq t\leq \Delta_P, t\leq \mathbf{r}\leq \Delta} F(\mathbf{r},t) =
F(\Delta,\Delta_P)$, which is true if both partial derivatives of $F$ are non-negative.\\
Taking the two partial derivatives of $F$:
$$\frac{\partial F}{\partial \mathbf{r}}(\mathbf{r},t) =  
 \frac{1}{E(c,\Delta)}
 \left(\frac{\partial E}{\partial \mathbf{r}}(c,\mathbf{r}) - \frac{\partial E}{\partial \mathbf{r}}(c,\mathbf{r}-t)\right)$$
$$\frac{\partial F}{\partial t}(\mathbf{r},t) =  
 \frac{ \frac{\partial E}{\partial \mathbf{r}} (c,\mathbf{r}-t)}{E(c,\Delta)}$$
In the proof for Theorem 1, we have proven that 
$\forall \mathbf{i}\leq 0, 0\leq \mathbf{r} < \Delta:$
$E(\mathbf{i},\mathbf{r}) \geq 0$, 
$\frac{\partial E}{\partial \mathbf{r}}(\mathbf{i},\mathbf{r}) \geq 0$ and 
$\frac{\partial^2 E}{\partial \mathbf{r}^2}(\mathbf{i},\mathbf{r})\geq 0$.\\
From 
$E(\mathbf{i},\mathbf{r}) \geq 0$, 
$\frac{\partial^2 E}{\partial \mathbf{r}^2}(\mathbf{i},\mathbf{r})\geq 0$
and $\mathbf{r} \geq \mathbf{r} -t$,
we conclude that $\frac{\partial F}{\partial k}(k,t)\geq 0$. \\
From 
$E(\mathbf{i},\mathbf{r}) \geq 0$ and
$\frac{\partial E}{\partial \mathbf{r}}(\mathbf{i},\mathbf{r}) \geq 0$,
we conclude that
$\frac{\partial F}{\partial t}(k,t)\geq 0$.\\
Hence,
$$|Q^+(c,\mathbf{r})-P_c^+(\hat{\mathbf{r}})| < \max_{0\leq t\leq \Delta_P, t\leq k\leq \Delta} F(k,t) = F(\Delta,\Delta_P) =  1- P_c^+(\Delta -\Delta_P)$$
Finally, the lemma follows from the fact that $P_c^+(\Delta -\Delta_P) \geq Q_{min}^+(\Delta -\Delta_P)$
\end{proof}
Lemma~\ref{lem5}, Lemma~\ref{lem6} and Lemma~\ref{lem7} are consecutively similar to 
Lemma~\ref{lem2}, Lemma~\ref{lem3} and Lemma~\ref{lem4}, but for $\Phi^-$.
The only difference is that  $Q^-(\mathbf{i},\mathbf{r})$ is increasing w.r.t $\mathbf{i}$
in the range $(-\infty,-1]$ and
the value of $Q^-(\mathbf{i},\mathbf{r})$ is inside the range
$(Q_{inf}^-(\mathbf{r}), Q^-_{max}(\mathbf{r}) ]$ where 
$Q^-_{inf}(\mathbf{r}) = \lim_{\mathbf{i} \to -\infty} Q^+(\mathbf{i},\mathbf{r}) $
and 
$Q_{max}^-(\mathbf{r})=Q^+ (-1,\mathbf{r})$.
\begin{lemma} \label{lem5}
$\forall \mathbf{i}\leq -1$, $0\leq \mathbf{r}<\Delta$:
$$\frac{\partial Q^-(\mathbf{i},\mathbf{r})}{\partial \mathbf{i}} \geq 0$$
\end{lemma}
\begin{proof}
Similar to Lemma~\ref{lem2}
\end{proof}

\begin{lemma} \label{lem6}
$\forall \mathbf{i}\leq -1,$ $c \leq -1$, $\mathbf{r}\in [0, \Delta)$:
$$|Q^-(\mathbf{i},\mathbf{r}) - Q^-(c,\mathbf{r})| < Q^-_{max}(\mathbf{r}^*) - Q^-_{inf}(\mathbf{r}^*)$$
where
$$Q^-_{inf}(\mathbf{r}) = 
\lim_{\mathbf{i} \to -\infty} Q^-(\mathbf{i},\mathbf{r})  =
\frac{2^{-\mathbf{r}} + \mathbf{r}\ln 2 - 1}{2^{-\Delta} + \Delta\ln 2 -1} $$
$$Q^-_{max}(\mathbf{r}) = Q^-(-1,\mathbf{r})
=\frac{ \ln (2-2^{-\mathbf{r}})-\mathbf{r}\ln 2 }
{ \ln (2-2^{-\Delta})-\Delta\ln 2 } $$
$$\mathbf{r}^* = \log_2\frac
{2X\ln X - X\ln(2X-1)}
{2X\ln X - 2X\ln(2X-1) + 2X -2 } \quad \text{with }
 X = 2^\Delta$$
\end{lemma}
\begin{proof}
The proof is similar to that of Lemma~\ref{lem3} but with some differences in function analysis.
See Appendix~\ref{ss:prooflem66} for proof details.
\end{proof}

\begin{lemma}\label{lem7}
Let  $\Delta_P$ be the distance of two adjacent index values of $P_c^-$
tables and $\hat{\mathbf{r}}$ be the rounding of $\mathbf{r}\in [0,\Delta)$ in modulo $\Delta_P$ , 
 i.e.,$\hat{\mathbf{r}} = \left\lceil \dfrac{x}{\Delta_P} \right\rceil \Delta_P$, then 
$$|Q^-(c,\mathbf{r})-P_c^-(\hat{\mathbf{r}})| \leq  1- Q_{inf}^-(\Delta -\Delta_P)  $$
\end{lemma}

\begin{proof}
Similar to  Lemma~\ref{lem4}
\end{proof}
Theorem~\ref{the2} derives the total error-bound of calculating $\Phi^+$ and $\Phi^-$
using the error-correction technique by accumulating the error-bound derived in 
Lemma~\ref{lem3} and Lemma~\ref{lem4} for $\Phi^+$ 
(similarly, Lemma~\ref{lem6} and Lemma~\ref{lem7} for $\Phi-$ ) with the fixed-point rounding errors.
\begin{theorembox}
\begin{theorem} \label{the2}
Let $\epsilon$ be the machine-epsilon of the fixed-point representation
of the LNS under consideration, and
the error-bounds of Interp-err of first-order Taylor approximation:
$E^+_M = E_\Delta^+(0), \; E^-_M = -E_\Delta^-(-1)$;
the error-bounds of Ratio-err:
$Q_R^+ = Q^+_{sup}(\mathbf{r}^*) - Q^+_{min}(\mathbf{r}^*), \;
Q_R^- = Q^-_{max}(\mathbf{r}^*) - Q^-_{inf}(\mathbf{r}^*)$;
the error-bounds of Index-err:
$Q_I^+ =  1- Q_{min}^+(\Delta -\Delta_P), \;
Q_I^- = 1- Q_{inf}^-(\Delta -\Delta_P)$,
then
$$| \Phi(x) -  \tilde{\Phi}_{EC}(x)| \leq (4+\Delta) \epsilon + E_M(Q_R + Q_I + \epsilon)$$
\end{theorem}
\end{theorembox}
\begin{proof}
Because $Q(\mathbf{i},\mathbf{r})=\frac{E_T(\mathbf{i}-\mathbf{r})}{ E_\Delta(\mathbf{i})}
= \frac{\Phi(x) -(\Phi(\mathbf{i}) -\mathbf{r}\Phi'(\mathbf{i})) }{ E_\Delta(\mathbf{i})}$,
we can write $\Phi(x)$ as:  
$\Phi(x) = \Phi(\mathbf{i}) -\mathbf{r}\Phi'(\mathbf{i}) + E_\Delta(\mathbf{i}) Q(\mathbf{i},\mathbf{r})$.
Together with Equation~\ref{eq:ECi}, the error of computing $\Phi(x)$ using the error-correction technique is:
$$ |\Phi(x) - \tilde{\Phi}_{EC}(x)| = 
|\Phi(\mathbf{i}) - \overline{\Phi}(\mathbf{i}) +
\text{rnd}(\mathbf{r}\overline{\Phi'}(\mathbf{i}))  -\mathbf{r}\Phi'(\mathbf{i})+
 E_\Delta(\mathbf{i}) Q(\mathbf{i},\mathbf{r}) - \text{rnd}(\overline{E_\Delta}(\mathbf{i})\overline{P_c}(\hat{\mathbf{r}}))| $$
We split $\Phi(\mathbf{i}) - \overline{\Phi}(\mathbf{i}) +
\text{rnd}(\mathbf{r}\overline{\Phi'}(\mathbf{i}))  -\mathbf{r}\Phi'(\mathbf{i})+
 E_\Delta(\mathbf{i}) Q(\mathbf{i},\mathbf{r}) - \text{rnd}(\overline{E_\Delta}(\mathbf{i})\overline{P_c}(\hat{\mathbf{r}})) $
into the sum of 8 terms $a_1, a_2,..., a_8$ and derive the error-bound of the absolute value of each term
as follows:
\begin{equation*}
\begin{aligned}
& a_1 =\Phi(\mathbf{i}) - \overline{\Phi}(\mathbf{i}) \Rightarrow |a_1| \leq \epsilon  \\
& a_2 = \text{rnd}(\mathbf{r}\overline{\Phi'}(\mathbf{i})) - \mathbf{r}\overline{\Phi'}(\mathbf{i}) \Rightarrow |a_2| \leq \epsilon \\
& a_3 =  \mathbf{r}\overline{\Phi'}(\mathbf{i})-  \mathbf{r}\Phi'(\mathbf{i}) \Rightarrow |a_3| \leq \mathbf{r}\epsilon \leq \Delta \epsilon\\
& a_4 = E_\Delta(\mathbf{i})Q(\mathbf{i},\mathbf{r}) - E_\Delta(\mathbf{i})Q(c,\mathbf{r}) \Rightarrow |a_4| \leq E_M Q_R\\
& a_5 = E_\Delta(\mathbf{i})Q(c,\mathbf{r}) - E_\Delta(\mathbf{i})P_c(\hat{\mathbf{r}})
\Rightarrow |a_5| \leq E_M Q_I\\
& a_6 = E_\Delta(\mathbf{i})P_c(\hat{\mathbf{r}}) - E_\Delta(\mathbf{i})\overline{P_c}(\hat{\mathbf{r}}) \Rightarrow |a_6| \leq E_M\epsilon \\
& a_7 = E_\Delta(\mathbf{i})\overline{P_c}(\hat{\mathbf{r}}) - \overline{E_\Delta}(x)\overline{P_c}(\hat{\mathbf{r}}) 
\Rightarrow |a_7| \leq  |\overline{P_c}(\hat{\mathbf{r}})| \epsilon \leq \epsilon \\
& a_8 = \overline{E_\Delta}(\mathbf{i})\overline{P_c}(\hat{\mathbf{r}}) - \text{rnd}(\overline{E_\Delta}(\mathbf{i})\overline{P_c}(\hat{\mathbf{r}})) \Rightarrow |a_8| \leq \epsilon 
\end{aligned}
\end{equation*}
\textit{Explanation:} $|a_1|$, $|a_2|$, and $|a_8|$ are just errors of fixed-point rounding, 
so they are at most $\epsilon$.
The bounds of $|a_3|$, $|a_7|$ are obtained from the fixed-point rounding error-bound $\epsilon$
together with the facts that $\mathbf{r} < \Delta$ and $\overline{P_c}(\hat{\mathbf{r}}) \leq 1$.
The bound of $E_\Delta$ is $E_M$, which is derived in Lemma~\ref{lem01} and Lemma~\ref{lem02}.
The bound of $|a_4|$ is obtained from 
the bound of the Ratio-err $|Q(\mathbf{i},\mathbf{r})- Q(c,\mathbf{r})|$, 
which is derived in Lemma~\ref{lem3} (for $\Phi^+$) and Lemma~\ref{lem6} (for $\Phi^-$).
The bound of $|a_5|$ is obtained from 
the bound of the Index-err $| Q(c,\mathbf{r}) - P_c(\hat{\mathbf{r}})|$,
which is derived in Lemma~\ref{lem4} (for $\Phi^+$) and Lemma~\ref{lem7} (for $\Phi^-$).
\\
Finally, we apply the absolute-value norm inequality $|\sum a_i| \leq \sum |a_i|$ to derive the total error-bound:
$$| \Phi(x) -  \hat{\Phi}_{EC}(x)| \leq (4+\Delta) \epsilon + E_M(Q_R + Q_I + \epsilon)$$

\end{proof}

Note that the error-bound derived in Theorem~\ref{the2} does not depend on the constant $c$.

\section{Rigorous error bound of the co-transformation technique}
\label{sec:cotrans}
This section mentions in full detail the derivation of the error-bound of computing $\Phi^-$
in the range $(-1,0)$ using the co-transformation technique, together with the formal mathematical proof.
The overview of the derivation was mentioned in Section~\ref{ss:CT}.
For a quick recall, the co-transformation technique's computation involves computing
$\Phi^-(x)$ by interpolation when $x$ contains errors.
In this case, the error-bound is derived by combining Lemma~\ref{lem31},  
which derives the error-bound of computing $\Phi^-(x)$ given the error-bound of $x$ 
assuming that computing $\Phi^-(x)$ is error-free,
with the error-bound of the interpolation technique,
which is derived in Theorem~\ref{theo1} (for first-order Taylor approximation) or
in Theorem~\ref{the2} (for the error-correction technique).
Finally, Theorem~\ref{theorem3} perform step-by-step
error-bound derivation for each of the three cases 
of the co-transformation technique's computation.
The total error-bound is the maximum of three error-bounds.

\begin{lemma}  \label{lem31}
For all $x,x^* \leq -1$ and $|x-x^*|\leq m$
$$|\Phi^-(x) - \Phi^-(x^*)| \leq \Phi^-(-1-m) - \Phi^-(-1)$$
\end{lemma}

\begin{proof}
Note that both the first and second derivatives of $\Phi^-$ are always negative:
$$\forall x: \quad (\Phi^-)'(x) = \frac{-2^x}{1-2^x} < 0 \quad \text{and} \quad
(\Phi^-)''(x) =\frac{-2^x\ln 2}{(1-2^x)^2} < 0$$
Without loss of generality, suppose that $x\geq x^*$, 
then $|\Phi^-(x) - \Phi^-(x^*)| = \Phi^-(x^*) - \Phi^-(x)$.\\
Let $t=x-x^*$ and $F(x,t) = \Phi^-(x-t) - \Phi^-(x)$, then
$0\leq t \leq m$ and 
$$F(x,t) = |\Phi^-(x) - \Phi^-(x^*)| \leq \max_{0\leq t\leq m, t\leq x\leq -1} F(x,t)$$
We will prove that $\max_{0\leq t\leq m, t\leq x\leq -1} F(x,t) =
F(-1,m)$, which is true if both partial derivatives of $F$ are non-negative.\\
Since $x \geq x-t$  and $(\Phi^-)''(x) <0 $:
$$\frac{\partial F}{\partial x} (x,t) 
= (\Phi^-)'(x-t) - (\Phi^-)' (x) \geq 0 $$
Since $(\Phi^-)'(x) <0 $:
$$\frac{\partial F}{\partial t} (x,t) 
= -(\Phi^-)'(x-t) \geq 0$$
Therefore, 
$\max_{ 0\leq t\leq m, t\leq x\leq -1} F(x,t) = F(-1,m) = \Phi^-(-1-m) - \Phi^-(-1)$.
\\
Hence, $|\Phi^-(k) - \Phi^-(k^*)|  \leq \Phi^-(-1-m) - \Phi^-(-1)$
\end{proof}

\begin{theorembox}
\begin{theorem} \label{theorem3}
Let $\epsilon$ be the 
machine-epsilon of the fixed-point representation 
of the LNS under consideration
and $E_{\Phi^-}$ be the error-bound of interpolating of $\Phi^-$
in the range $(-\infty,1]$.
The error-bound of computing $\Phi^-(x)$
when $x \in (-1,0)$ using the co-transformation technique is:
$$\max(\epsilon + \Phi^-(-1-2\epsilon) - \Phi^-(-1) + E_{\Phi^-}\;\;,\;\;
\epsilon + \Phi^-(-1-E_{k_2}) - \Phi^-(-1) + E_{\Phi^-})$$
where
$$E_{k_2} = 2\epsilon + \Phi^-(-1-2\epsilon) - \Phi^-(-1) + E_{\Phi^-}$$ 
\end{theorem}
\end{theorembox}

\begin{proof}
We derive the error-bound for each of the three cases of the co-transformation technique.\\
\vspace{1ex}
\noindent$\bullet\; ${\bf Case 1:\/}  $x \in [-\Delta_a, 0)$: $\Phi^-(x)$ is indexed directly from $T_a$, so the error-bound is $\epsilon$.

\vspace{1ex}
\noindent$\bullet\; ${\bf Case 2:\/} $x \in [-\Delta_b, -\Delta_a)$:
$r_b$ and  $r_a$ are error-free 
because their computations involve only operations that are error-free in fixed-point arithmetic,
we consecutively derive the error-bound of computing $k$, $\Phi^-(k)$ and finally $\Phi^-(x)$.
Let $\tilde{k}$ and $\tilde{\Phi^-}$ be the results of the
computations of $k$ and $\Phi^-$ (with fixed-point rounding and interpolation errors).
From the formula $ k = x-\Phi^-(r_b)+\Phi^-(r_a)$, 
the error of computing $k$ consists of 
the two fixed-point rounding errors of $\Phi^-(r_a)$ and $\Phi^-(r_b)$,
so its error-bound is $2\epsilon$, (i.e. $|k-\tilde{k}| \leq 2\epsilon$).
From the error-bound of computing $k$ and Lemma~\ref{lem31}, 
we get $|\Phi^-(k) - \Phi^-(\tilde{k})| \leq \Phi^-(-1-2\epsilon) - \Phi^-(-1)$. 
Next, because $\tilde{\Phi^-}(\tilde{k})$ is computed by interpolation,
$|\Phi^-(\tilde{k})-\tilde{\Phi^-}(\tilde{k})| \leq  E_{\Phi^-}$,
we apply the absolute-value norm inequality to derive the error-bound of computing $\Phi^-(k)$ :
$$|\Phi^-(k) - \tilde{\Phi^-}(\tilde{k})|
\leq |\Phi^-(k) - \Phi^-(\tilde{k})|+ 
|\Phi^-(\tilde{k})-\tilde{\Phi^-}(\tilde{k})| \leq \Phi^-(-1-2\epsilon) - \Phi^-(-1) + E_{\Phi^-}$$
Finally, from Equation~\ref{eq:CT2}, we accumulate 
the error-bound of computing $\Phi^-(k)$ with 
the fixed-point rounding error-bound of $\Phi^-(r_b)$, which is $\epsilon$,
to get the error-bound of computing $\Phi^-(x)$ in Case 2: 
$$\epsilon + \Phi^-(-1-2\epsilon) - \Phi^-(-1) + E_{\Phi^-}$$
\vspace{1ex}
\noindent$\bullet\; ${\bf Case 3:\/} $x \in (-1, -\Delta_b):$
Similar to Case 2, all the terms $r_c$, $r_{ab}$, $r_b$ and $r_a$ are error-free,
we consecutively derive the error-bound of computing $k_1$, $\Phi^-(k_1)$,
$k_2$, $\Phi^-(k_2)$,
and finally $\Phi^-(x)$.
Let $\tilde{k_1}$, $\tilde{k_2}$ and
$\tilde{\Phi^-}$
be the actual results of
computation of $k_1$, $k_2$ and
$\Phi^-$.
Because of the similarity in the formula of $k_1$ and that of
$k$ in Case 2, the error-bound of computing $k_1$ and $\Phi^-(k_1)$ 
is the same as that of computing $k$ and $\Phi^-(k)$ in Case 2,
which are $2\epsilon$ and 
$\Phi^-(-1-2\epsilon) - \Phi^-(-1) + E_{\Phi^-}$.
From the formula $k_2 = x+\Phi^-(r_b) + \Phi^-(k_1)-\Phi^-(r_c)$,
we derive the error-bound of computing $k_2$ by accumulating
the error-bound of computing $\Phi^-(k_1)$ with
the fixed-point rounding error-bounds of 
$\Phi^-(r_b)$ and  $\Phi^-(r_c)$ (both are $\epsilon$).
Let $E_{k_2}$ be the error-bound of computing $k_2$, then
$$E_{k_2} = 2\epsilon + \Phi^-(-1-2\epsilon) - \Phi^-(-1) + E_{\Phi^-}$$
Similar to how we derive the error-bound of computing $\Phi^-(k_1)$, 
the error-bound of computing $\Phi^-(k_2)$ is:
$$|\Phi^-(k_2) - \tilde{\Phi^-}(\tilde{k_2})|
\leq\Phi^-(-1-E_{k_2}) - \Phi^-(-1) + E_{\Phi^-}$$
Finally, from Equation~\ref{eq:CT3}, we accumulate 
the error-bound of computing $\Phi^-(k_2)$ with 
the fixed-point rounding error-bound of $\Phi^-(r_c)$, which is $\epsilon$,
to get the error-bound of computing $\Phi^-(x)$ in Case 3: 
$$\epsilon + \Phi^-(-1-E_{k_2}) - \Phi^-(-1) + E_{\Phi^-}$$
\\
Hence, the error-bound of computing $\Phi^-(x)$
when $-1<x<0$ with co-transformation technique is
the maximum of the error-bounds of {\bf  Case 2} and 
{\bf  Case 3}, which is:
$$\max(\epsilon + \Phi^-(-1-2\epsilon) - \Phi^-(-1) + E_{\Phi^-}\;\;, \;\;
\epsilon + \Phi^-(-1-E_{k_2}) - \Phi^-(-1) + E_{\Phi^-})$$
where
$$E_{k_2} = 2\epsilon + \Phi^-(-1-2\epsilon) - \Phi^-(-1) + E_{\Phi^-}$$ 
\end{proof}

\section{Numerical experiments}
 
\label{sec:numerical-expts}


In this section, we perform two experiments.
The first experiment (Section~\ref{ss:verifyexp}) use numerical testing 
to verify the correctness of the error-bounds' formulas
of computing $\Phi^+/\Phi^-$ using 
first-order Taylor approximation (derived in Theorems~\ref{theo1}),
the error-correction technique (derived in Theorems~\ref{the2}),
and the co-transformation technique (derived in Theorems~\ref{theorem3}).
For each technique, we test the error-bounds over different combination of values
of the parameters: $\epsilon, \Delta, \Delta_P, \Delta_a $ and $\Delta_b$.
For a quick recall, $\epsilon$ is 
the machine-epsilon of the fixed-point representation 
of the LNS under consideration;
$\Delta$
are the separation of adjacent values in the look-up tables of $\Phi$ and $\Phi'$; 
$\Delta_P, \Delta_a $ and $\Delta_b$ are those of tables $P_c$, $T_a$, and $T_b$.
The second experiment (Section~\ref{ss:parameterimpact}) 
uses the error-bounds' formulas to measure the impact
of the interpolation techniques and the parameters on the magnitude of the 
relative error-bounds.
The result of the experiment provides insight into the problem of selecting 
LNS designs that match some desired accuracy and hardware limitations.

\subsection{Numerical verification of error-bound}
\label{ss:verifyexp}
The derived error-bound formulas contain non-polynomial functions $\Phi^+$ and $\Phi^-$, 
which cannot be handled by current automated theorem provers, and we chose an
expedient way to check the bounds, namely through empirical testing.
%
%
In the scope of our work, we perform the following experiment to verify the derived error-bound formulas with numerical values.


%
In total, we have derived the error-bound formulas for 6 cases of interpolation:
\begin{compactenum}
    \item Computing $\Phi^+(x)$ for $x\in (-\infty, 0]$ using first-order Taylor approximation (Theorem~\ref{theo1}).
    \item Computing $\Phi^+$ for $x\in (-\infty, 0]$ using the error-correction technique (Theorem~\ref{the2}).
    \item Computing $\Phi^-(x)$ for $x\in (-\infty, -1]$ using first-order Taylor approximation (Theorem~\ref{theo1}). 
    \item Computing $\Phi^-(x)$ for $x\in (-\infty, -1]$ using the error-correction technique (Theorem~\ref{the2}).
    \item Computing $\Phi^-(x)$ for $x\in (-1, 0)$ using the co-transformation technique 
      combined with first-order Taylor approximation (Theorem~\ref{theorem3} with $E_{\Phi^-}$ computed by Theorem~\ref{theo1}).
    \item Computing $\Phi^-(x)$ for $x\in (-1, 0)$ using the co-transformation technique 
      combined with error-correction technique (Theorem~\ref{theorem3} with $E_{\Phi^-}$ computed by Theorem~\ref{the2}). 
\end{compactenum}
The computation of each case involves a set of parameters:

\begin{compactitem}
    \item Computing $\Phi^+$/$\Phi^-$ using first-order Taylor involves  $\epsilon$ and $\Delta$
    \item Computing $\Phi^+$/$\Phi^-$ using the error-correction technique involves  
    $\epsilon$, $\Delta$ and $\Delta_P$
    \item Computing $\Phi^-$ using the co-transformation technique involves  the parameters of the interpolation technique (first-order Taylor approximation or the error-correction technique) 
    together with the two parameters: $\Delta_a$ and $\Delta_b$.
\end{compactitem}

\paragraph{Choice of Experimental Parameters} Note that although the computation of the error-correction technique involves the constant $c$,
in our experiments, we do not consider $c$ as a parameter for the following reasons:
\begin{compactitem}
    \item In actual LNS design, $c$ is not selected arbitrarily, but rather carefully to minimize the error of computing $\Phi$ (see \cite{AEM}). Finding the optimized value of $c$ for each test case is beyond the scope of this experiment.
    
    \item For the purpose of verifying the validity of the error-bounds, if the error-bounds are valid on an
    {\em arbitrary} value of $c$, they are guaranteed to be valid on the value of $c$ that minimizes the error of computing $\Phi$.
     When evaluating a bound that
     uses error-correction techniques, 
     {\em we set $c=-4$, which is the optimized value
of $c$ in European Logarithm Microprocessor}.~\cite{AEM}.
    
    \item The error-bound of the error-correction technique {\em does not depend} on $c$.
\end{compactitem}

We perform $76$ test cases, which cover all 6 cases of interpolation with various combination
of values of the parameters.
The details of each test case are listed in Tables~\ref{ftaddsubtab},~\ref{ecaddsubtab},~\ref{cotranstesttab}.
For each test case, we pre-specify a sample set, an interpolation method, and the values of the parameters,
then accordingly compute $\Phi^+$/$\Phi^-$ for all values in the sample set.
The error of computation is estimated by the difference between the computed results with the values
of $\Phi^+$/$\Phi^-$ computed directly in double-precision by Numpy library.
The values of parameters are selected so that the size of all look-up tables are at most $2^{10}$,
which is suitable for them to be stored in ROM.
During the experiment's design, we gradually increased the size and density of the sample sets 
such that they included enough points that represent extreme cases of the bounds' tightness.

\begin{table}[H]
\begin{tabular}{|l|l|l|l|}
\hline
Name             & $\epsilon$                 & $\Delta$ & Sample set                                                               \\ \hline
FT-Add1, FT-Sub1 & \multirow{3}{*}{$2^{-8}$}  & $2^{-3}$ & \multirow{3}{*}{\shortstack{For FT-Adds: $\{-k2^{-8}:\; k = 0, 1, ..., 3\times 2^8\}$\\For FT-Subs: $\{-k2^{-8}-1:\; k = 0, 1, ..., 3\times 2^8\}$}} \\
FT-Add2, FT-Sub2 &                            & $2^{-4}$ &                                                                          \\
FT-Add3, FT-Sub3 &                            & $2^{-5}$ &                                                                          \\ \cline{2-4}
FT-Add4, FT-Sub4 & \multirow{3}{*}{$2^{-16}$} & $2^{-4}$ & \multirow{3}{*}{\shortstack{For FT-Adds: $\{-k2^{-16}:\; k = 0, 1, ..., 3\times 2^{16}\}$\\For FT-Subs: $\{-k2^{-8}-1:\; k = 0, 1, ..., 3\times 2^{16}\}$}}                                                        \\
FT-Add5, FT-Sub5 &                            & $2^{-6}$ &                                                                          \\
FT-Add6, FT-Sub6 &                            & $2^{-8}$ &                                                                          \\ \cline{2-4}
FT-Add7, FT-Sub7 & \multirow{3}{*}{$2^{-32}$} & $2^{-4}$ & \multirow{3}{*}{\shortstack{For FT-Adds: $\{-k2^{-32}:\; k = 0, 1, ..., 3\times 2^{32}\}$\\For FT-Subs: $\{-k2^{-8}-1:\; k = 0, 1, ..., 3\times 2^{32}\}$}}                                                        \\
FT-Add8, FT-Sub8 &                            & $2^{-6}$ &                                                                          \\
FT-Add9, FT-Sub9 &                            & $2^{-8}$ &                                                                          \\ \hline
\end{tabular}
\caption{18 test cases for computing $\Phi^+$/$\Phi^-$
using first-order Taylor approximation.}
\label{ftaddsubtab}
\end{table}

\begin{table}[H]
\begin{tabular}{|l|l|l|l|l|}
\hline
Name               & $\epsilon$                 & $\Delta$                  & $\Delta_P$ & Sample set                                                                                                                                              \\ \hline
EC-Add1, EC-Sub1   & \multirow{5}{*}{$2^{-8}$}  & \multirow{2}{*}{$2^{-3}$} & $2^{-6}$   & \multirow{5}{*}{\shortstack{For EC-Adds: $\{-k2^{-8}:\; k = 0, 1, ..., 3\times 2^8\}$\\For EC-Subs: $\{-k2^{-8}-1:\; k = 0, 1, ..., 3\times 2^8\}$}}        \\
EC-Add2, EC-Sub2   &                            &                           & $2^{-7}$   &                                                                                                                                                         \\ \cline{3-4}
EC-Add3, EC-Sub3   &                            & \multirow{2}{*}{$2^{-4}$} & $2^{-7}$   &                                                                                                                                                         \\
EC-Add4, EC-Sub4   &                            &                           & $2^{-8}$   &                                                                                                                                                         \\ \cline{3-4}
EC-Add5, EC-Sub5   &                            & $2^{-5}$                  & $2^{-8}$   &                                                                                                                                                         \\ \cline{2-5} 
EC-Add6, EC-Sub6   & \multirow{6}{*}{$2^{-16}$} & \multirow{2}{*}{$2^{-4}$} & $2^{-7}$   & \multirow{6}{*}{\shortstack{For EC-Adds: $\{-k2^{-16}:\; k = 0, 1, ..., 3\times 2^{16}\}$\\For EC-Subs: $\{-k2^{-16}-1:\; k = 0, 1, ..., 3\times 2^{16}\}$}} \\
EC-Add7, EC-Sub7   &                            &                           & $2^{-8}$   &                                                                                                                                                         \\ \cline{3-4}
EC-Add8, EC-Sub8   &                            & \multirow{2}{*}{$2^{-6}$} & $2^{-9}$   &                                                                                                                                                         \\
EC-Add9, EC-Sub9   &                            &                           & $2^{-10}$  &                                                                                                                                                         \\ \cline{3-4}
EC-Add10, EC-Sub10 &                            & \multirow{2}{*}{$2^{-8}$} & $2^{-11}$  &                                                                                                                                                         \\
EC-Add11, EC-Sub11 &                            &                           & $2^{-12}$  &                                                                                                                                                         \\ \cline{2-5} 
EC-Add12, EC-Sub12 & \multirow{6}{*}{$2^{-32}$} & \multirow{2}{*}{$2^{-4}$} & $2^{-7}$   & \multirow{6}{*}{\shortstack{For EC-Adds: $\{-k2^{-16}:\; k = 0, 1, ..., 3\times 2^{16}\}$\\For EC-Subs: $\{-k2^{-16}-1:\; k = 0, 1, ..., 3\times 2^{16}\}$}} \\
EC-Add13, EC-Sub13 &                            &                           & $2^{-8}$   &                                                                                                                                                         \\ \cline{3-4}
EC-Add14, EC-Sub14 &                            & \multirow{2}{*}{$2^{-6}$} & $2^{-9}$   &                                                                                                                                                         \\
EC-Add15, EC-Sub15 &                            &                           & $2^{-10}$  &                                                                                                                                                         \\ \cline{3-4}
EC-Add16, EC-Sub16 &                            & \multirow{2}{*}{$2^{-8}$} & $2^{-11}$  &                                                                                                                                                         \\
EC-Add17, EC-Sub17 &                            &                           & $2^{-12}$  &                                                                                                                                                         \\ \hline
\end{tabular}
\caption{34 Test cases for computing $\Phi^+$/$\Phi^-$ using error-correction technique.}
\label{ecaddsubtab}
\end{table}

\begin{table}[H]
\begin{tabular}{|l|l|l|l|l|l|}
\hline
Name      & Interpolation                        & $\epsilon$                 & $(\Delta_a, \Delta_b)$               & $\Delta$ & Sample set                                                   \\ \hline
Cotrans1  & \multirow{12}{*}{First-order Taylor} & \multirow{4}{*}{$2^{-8}$}  & \multirow{2}{*}{$(2^{-6},2^{-3})$}   & $2^{-3}$ & \multirow{4}{*}{$\{-k2^{-8}:\; k = 0, 1, ..., 2^8-1\}$}   \\ \cline{5-5}
Cotrans2  &                                      &                            &                                      & $2^{-4}$ &                                                              \\ \cline{4-5}
Cotrans3  &                                      &                            & \multirow{2}{*}{$(2^{-5},2^{-2})$}   & $2^{-3}$ &                                                              \\ \cline{5-5}
Cotrans4  &                                      &                            &                                      & $2^{-4}$ &                                                              \\ \cline{3-6} 
Cotrans5  &                                      & \multirow{4}{*}{$2^{-16}$} & \multirow{2}{*}{$(2^{-12},2^{-6})$}  & $2^{-4}$ & \multirow{8}{*}{$\{-k2^{-16}:\; k = 0, 1, ..., 2^{16}-1\}$} \\ \cline{5-5}
Cotrans6  &                                      &                            &                                      & $2^{-6}$ &                                                              \\ \cline{4-5}
Cotrans7  &                                      &                            & \multirow{2}{*}{$(2^{-10},2^{-5})$}  & $2^{-4}$ &                                                              \\ \cline{5-5}
Cotrans8  &                                      &                            &                                      & $2^{-6}$ &                                                              \\ \cline{3-5}
Cotrans9  &                                      & \multirow{4}{*}{$2^{-32}$} & \multirow{2}{*}{$(2^{-22},2^{-11})$} & $2^{-4}$ &                                                              \\ \cline{5-5}
Cotrans10 &                                      &                            &                                      & $2^{-6}$ &                                                              \\ \cline{4-5}
Cotrans11 &                                      &                            & \multirow{2}{*}{$(2^{-20},2^{-10})$} & $2^{-4}$ &                                                              \\ \cline{5-5}
Cotrans12 &                                      &                            &                                      & $2^{-6}$ &                                                              \\ \cline{2-6} 
Cotrans13 & \multirow{12}{*}{Error-correction}   & \multirow{4}{*}{$2^{-8}$}  & \multirow{2}{*}{$(2^{-6},2^{-3})$}   & $2^{-3}$ & \multirow{4}{*}{$\{-k2^{-8}:\; k = 0, 1, ..., 2^8-1\}$}   \\ \cline{5-5}
Cotrans14 &                                      &                            &                                      & $2^{-4}$ &                                                              \\ \cline{4-5}
Cotrans15 &                                      &                            & \multirow{2}{*}{$(2^{-5},2^{-2})$}   & $2^{-3}$ &                                                              \\ \cline{5-5}
Cotrans16 &                                      &                            &                                      & $2^{-4}$ &                                                              \\ \cline{3-6} 
Cotrans17 &                                      & \multirow{4}{*}{$2^{-16}$} & \multirow{2}{*}{$(2^{-12},2^{-6})$}  & $2^{-4}$ & \multirow{8}{*}{$\{-k2^{-16}:\; k = 0, 1, ..., 2^{16}-1\}$} \\ \cline{5-5}
Cotrans18 &                                      &                            &                                      & $2^{-6}$ &                                                              \\ \cline{4-5}
Cotrans19 &                                      &                            & \multirow{2}{*}{$(2^{-10},2^{-5})$}  & $2^{-4}$ &                                                              \\ \cline{5-5}
Cotrans20 &                                      &                            &                                      & $2^{-6}$ &                                                              \\ \cline{3-5}
Cotrans21 &                                      & \multirow{4}{*}{$2^{-32}$} & \multirow{2}{*}{$(2^{-22},2^{-11})$} & $2^{-4}$ &                                                              \\ \cline{5-5}
Cotrans22 &                                      &                            &                                      & $2^{-6}$ &                                                              \\ \cline{4-5}
Cotrans23 &                                      &                            & \multirow{2}{*}{$(2^{-20},2^{-10})$} & $2^{-4}$ &                                                              \\ \cline{5-5}
Cotrans24 &                                      &                            &                                      & $2^{-6}$ &                                                              \\ \hline
\end{tabular}
\caption{24 test cases for computing $\Phi^-$ in the range $(-1,0)$ using co-transformation technique. For the error-correction technique, we set $\Delta_P = 2^{-3}\Delta $.}
\label{cotranstesttab}
\end{table}

\begin{figure}[H]
\includegraphics[scale=0.8]{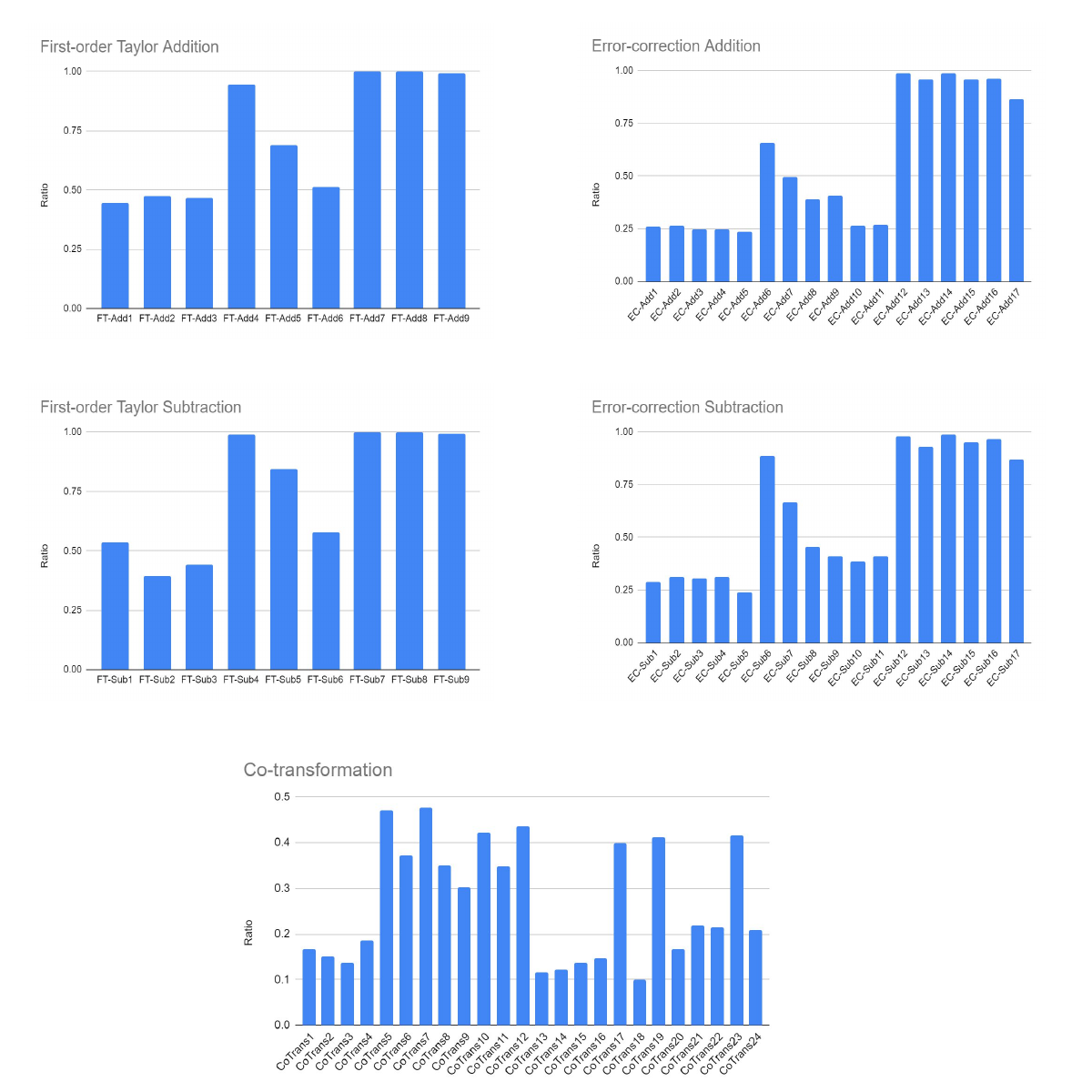}
  \caption{The ratio of the maximum error of test cases to the error-bounds computed by the derived formulas.
  Each column represents a test case in Tables~\ref{ftaddsubtab},~\ref{ecaddsubtab},~\ref{cotranstesttab}.
  }
  
  \label{fig:emaxebound}
\end{figure}

\begin{figure}[H]
\includegraphics[scale=0.63]{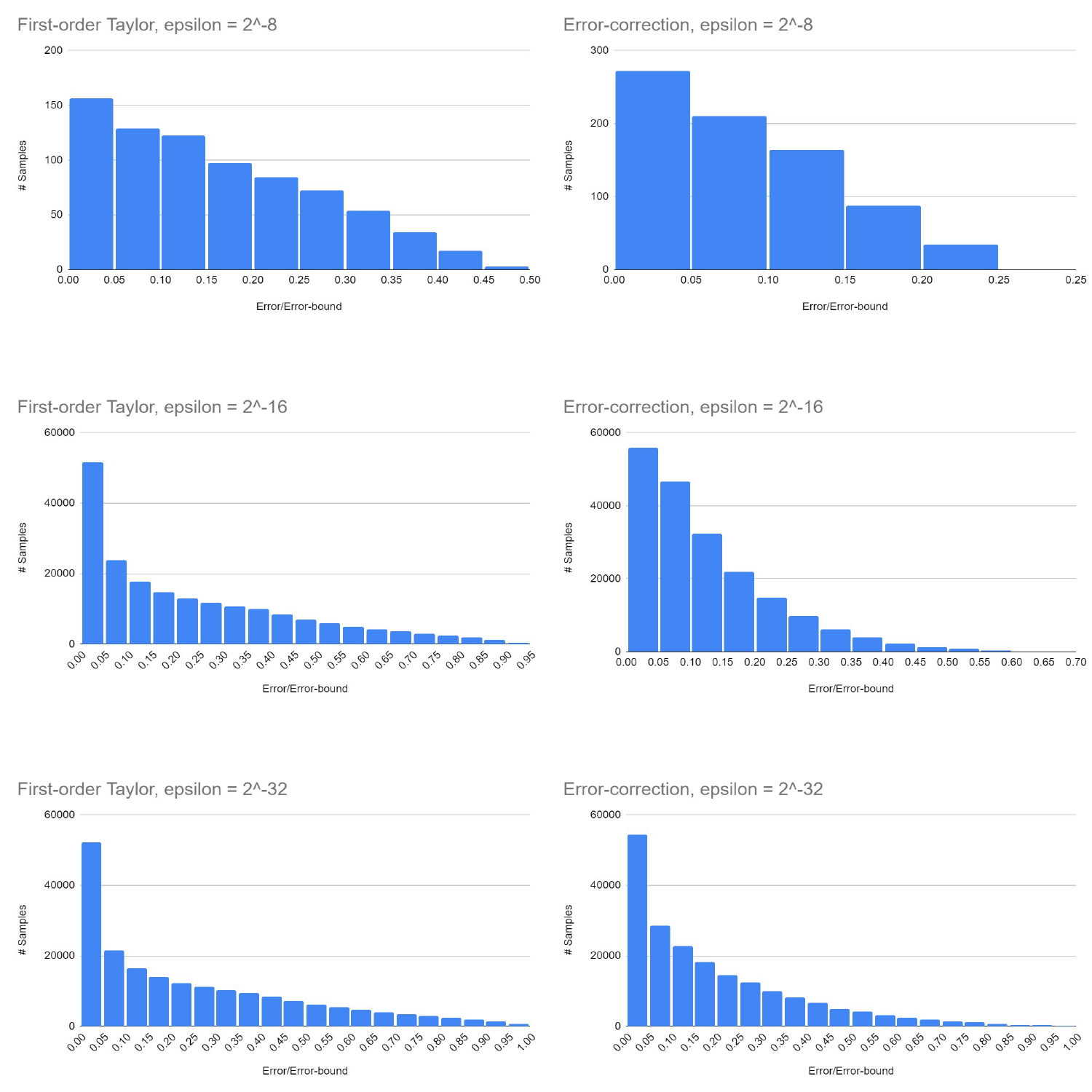}
  \caption{The histograms of Error/Error-bound computing addition by first-order Taylor approximation and error-correction technique with different values of $\epsilon$ where $\Delta$ and $\Delta_P$ are fixed to be $2^{-4}$
  and $2^{-7}$}
  \label{fig:addhistogram}
\end{figure}

\begin{figure}[H]
\includegraphics[scale=0.63]{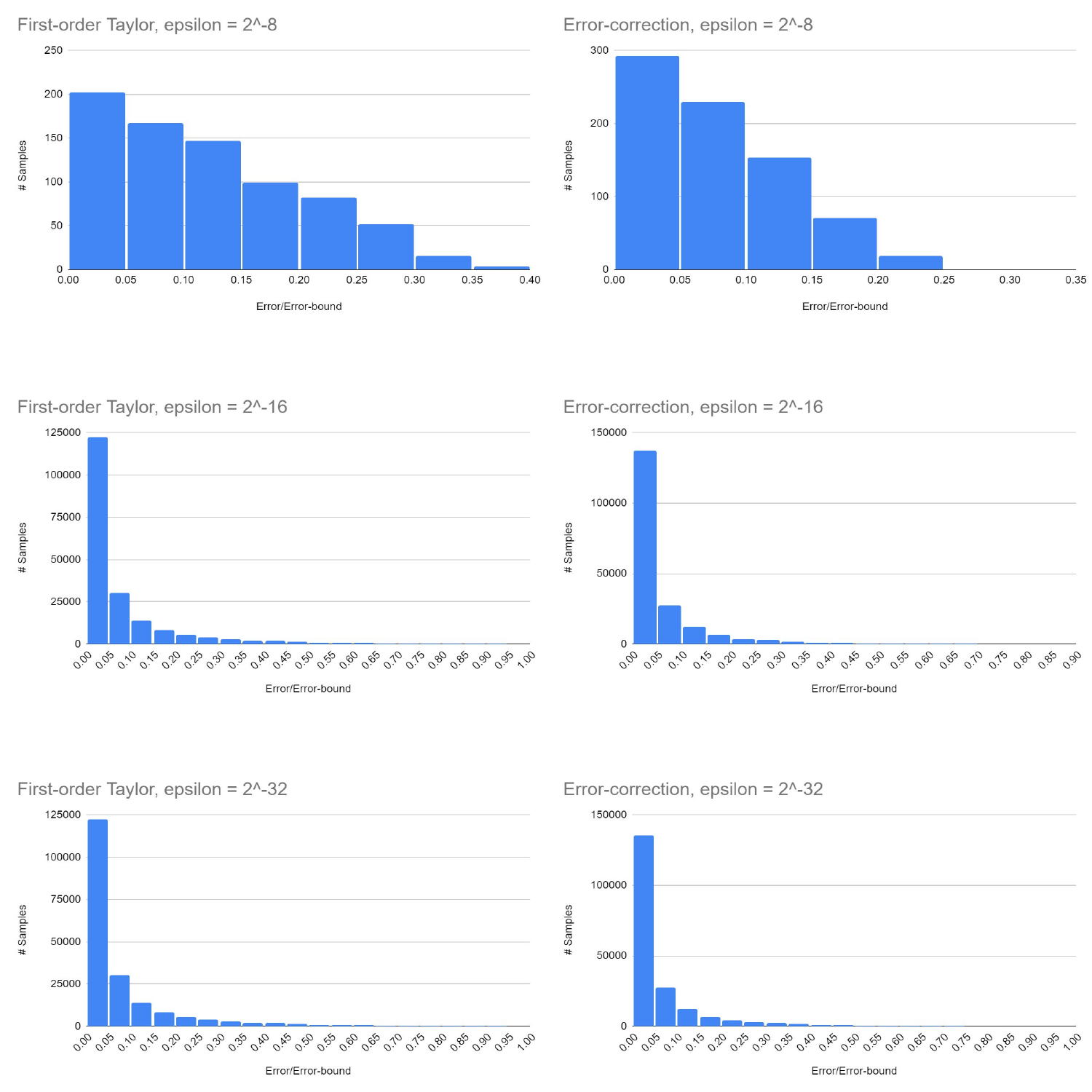}
  \caption{The histograms of Error/Error-bound computing subtraction by first-order Taylor approximation and error-correction technique with different values of $\epsilon$ where $\Delta$ and $\Delta_P$ are fixed to be $2^{-4}$ and $2^{-7}$}
  \label{fig:subhistogram}
\end{figure}

\begin{figure}[H]
\includegraphics[scale=0.63]{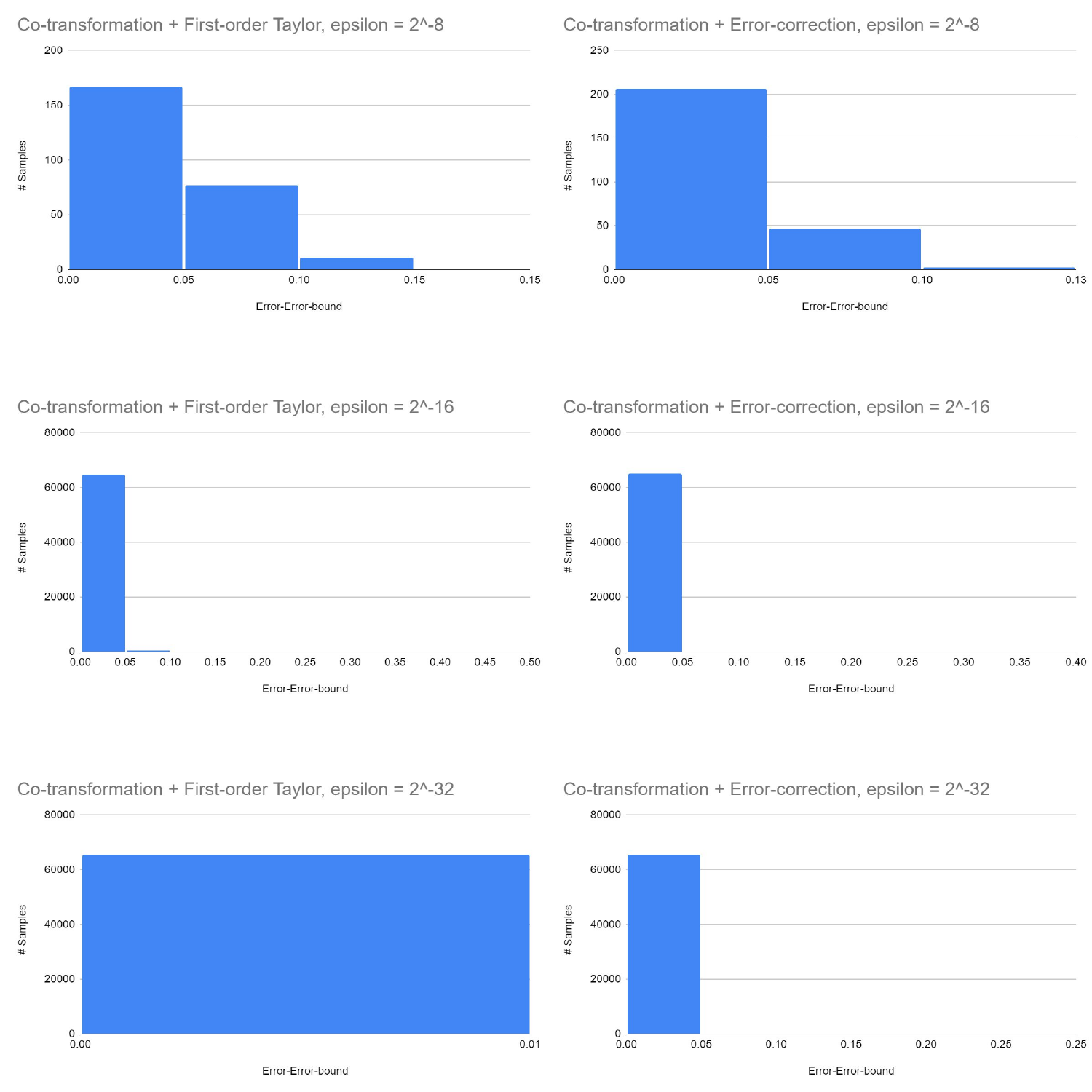}
  \caption{The histograms of Error/Error-bound computing subtraction by co-transformation technique with different values of $\epsilon$ where $\Delta$ and $\Delta_P$ are fixed to be $2^{-4}$ and $2^{-7}$}
  \label{fig:cotranshistogram}
\end{figure}

Figure~\ref{fig:emaxebound} shows the ratio of the maximum error of each test case to the derived error-bound.
All the ratios are below $1.0$, which strongly suggests that the error-bound is rigorous.
Furthermore, the graph also indicates that the higher precision of the fix-point representation
for implementing the LNS (the smaller $\epsilon$), the tighter the error-bounds are.
Figure~\ref{fig:addhistogram} and ~\ref{fig:subhistogram}
plot the histograms of
the ratios of the errors of the testing samples to the error-bound of
computing addition and subtraction using first-order Taylor approximation and the error-correction technique 
with fixed values of $\Delta = 2^{-4}$, $\Delta_P = 2^{-7}$ and three values of $\epsilon:$
$ 2^{-8}, 2^{-16}$ and $2^{-32}$.
In other words, Figure~\ref{fig:addhistogram} plots the histograms of test cases: 
FT-Add2, FT-Add4, FT-Add7, EC-Add3, EC-Add6, EC-Add12; 
and Figure~\ref{fig:subhistogram} plots the histograms of test cases: 
FT-Sub2, FT-Sub4, FT-Sub7, EC-Sub3, EC-Sub6, EC-Sub12.
Figure~\ref{fig:cotranshistogram} 
plots similar histograms for the co-transformation technique combined with 
first-order Taylor approximation (test cases: Cotrans2, Cotrans5, Cotrans9) 
and the error-correction technique (test cases: Cotrans14, Cotrans17, Cotrans21).
Those histograms indicate that although the maximum error is close to the error-bound
in some cases, the ratios of most errors of testing samples to the error-bound
are very small.\\
In summary, our testing results establish the reliability of our error bounds
---a valuable result that avoids over-provisioning in terms of hardware/software designs.

\subsection{Magnitude of error-bound over parameters}
\label{ss:parameterimpact}
One important application of deriving rigorous error-bound formulas is 
knowing how to tune the LNS parameters in order to achieve the desired error-bound
for arithmetic operators.
Note that although the computation of the co-transformation technique
requires the two parameters $\Delta_a$ and $\Delta_b$,
our derived error-bound does not depend on those parameters.
In this section, we only consider the three parameters which effect
the magnitude of the error-bounds: $\epsilon$, $\Delta$ and $\Delta_p$.

\paragraph{Table Sizes}
Note that we can easily compute the sizes of look-up tables 
from $\epsilon$, $\Delta$, $\Delta_P$ and $l$, which is the number of bits used to 
represent the exponent of LNS:
\begin{compactitem}
    \item The size of the look-up tables for $\Phi$, $\Phi'$ and $E_\Delta$ is 
    $2^{l-\log_2 \Delta + \log_2 \epsilon}$
    \item The size of the look-up tables for $P_c$ is $\Delta/\Delta_P$
\end{compactitem}
The reader may find specific table sizes discussed in other LNS publications (such a discussion is outside the focus of this paper).

The following experiment provides insight into that problem by 
exploring the impact of the interpolation methods and their parameters 
on the magnitude of the error-bound.
We perform this experiment on the bounds of the relative error (relative-error-bounds)
of LNS's addition/subtraction  
instead of the error-bound of computing $\Phi^+$/$\Phi^-$,
because the relative-error-bound is more widely used in 
high-dynamic-range, non-uniform-distributed number systems 
(such as LNS and floating-point)
and can be easily computed from the error-bound of computing $\Phi^+$/$\Phi^-$
by the following Corollary:
\begin{corollary}\label{cl1}
Let $\mathit{U}$ be the the absolute error-bound of computing $\Phi^+$/$\Phi^-$,
then the relative error-bound of LNS's addition/subtraction is $2^\mathit{U}$
\end{corollary}
\begin{proof}
From the equations of LNS's addition/subtraction: $\log_2(2^i \pm 2^j) = i + \Phi^{\pm}(j-i)$,
let $\mathit{err}$ be the the absolute error of computing $\Phi^+$/$\Phi^-$. 
Because the fixed-point addition with $i$ is error-free,
$\mathit{err}$ is also that of the absolute error of the
fixed-point representation of the result of addition/subtraction ( $\log_2(2^i \pm 2^j)$),
and $2^\mathit{err}$ is the relative error of the real result.
The Corollary's result follows from the additional fact that $2^x$ is a monotonical
increasing function.
\end{proof}
Specifically, the purpose of this experiment is to compare:
\begin{enumerate}
    \item[-] The difference between the relative-error-bounds when applying
first-order Taylor approximation versus the error-correction technique.
    \item[-] The impact of $\Delta$ and $\Delta_p$ on the magnitude of the relative-error-bounds.    
    \item[-] The difference between the relative-error-bounds when we apply the
     co-transformation technique with first-order Taylor approximation and 
     with the error-correction technique.
\end{enumerate}
In this experiment, we compute the relative-error-bounds of addition and subtraction 
for 6 cases of $\Phi^+$/$\Phi^-$'s interpolation, 
which are mentioned at the beginning of Section~\ref{sec:numerical-expts}.1.
For each case, we compute the relative-error-bound for each value of $\Delta$ in the set: $\{2^{-3}, 2^{-4},..., 2^{-10},\}$.
When the error-correction technique is applied, with each value of $\Delta$,
we compute the relative-error-bound for each value of $\Delta_P$
in the set  $\{2^{-4}\Delta ,   2^{-8}\Delta \}$.
We choose $\epsilon=2^{-24}$, which is the same as the machine epsilon of Floating-point 32.
{ These constants were chosen in the European Logarithmic Microprocessor~\cite{EM}. In MADAM, $\epsilon$ is much coarser, appearing to fall, according to~\cite{madam}, in the range $[2^{-10},2^{-8}]$, storing the numbers as integers but with a base of $2^{-10}$ with no error-correction employed.}
\begin{figure}[H]
\includegraphics[scale=0.67]{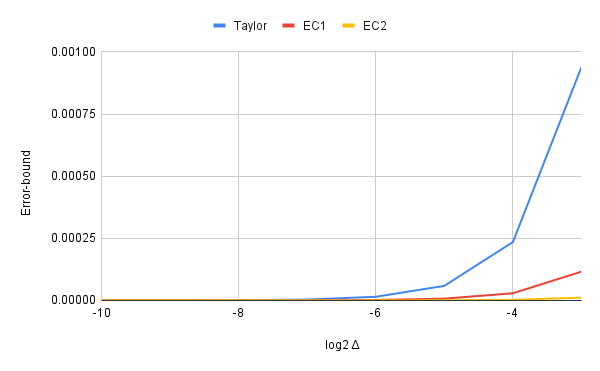}
  \caption{Relative-error-bound of
  addition
  (of numbers falling
  within the whole domain
  supported by the underlying fixed-point representation)
  using first-order Taylor approximation (Taylor),
  error-correction with $\Delta_p=2^{-4}\Delta$ (EC1) and with $\Delta_p=2^{-8}\Delta$ (EC2)}
  \label{fig:Phi+chart}
\end{figure}
\begin{figure}[H]
\includegraphics[scale=0.6]{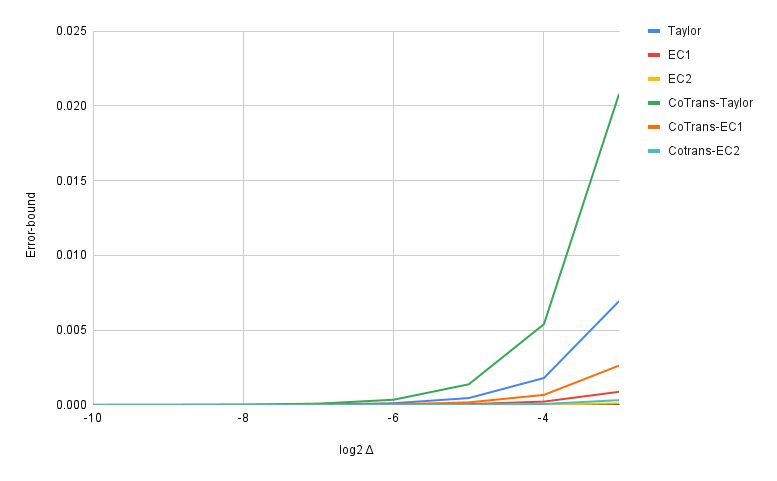}
  \caption{Relative-error-bound of subtraction using first-order Taylor approximation (Taylor),
  error-correction with $\Delta_p=2^{-4}\Delta$ (EC1) and with $\Delta_p=2^{-8}\Delta$ (EC2) 
  with and without co-transformation (Cotrans) }
  \label{fig:Phi-chart}
\end{figure}
Figures~\ref{fig:Phi+chart} and \ref{fig:Phi-chart}
plot of the relative error-bounds of LNS's addition and subtraction over $\log_2 \Delta$
The key takeaways are the following:
\begin{compactitem}
\item[-] The greater the value of $\Delta$ is, the more significant the difference
between the relative-error-bounds of
using first-order Taylor approximation and those of using the error-correction technique,
and the more significant difference between the relative-error-bounds of using an error-correction technique
with different values of $\Delta_P$ (see Figure~\ref{fig:Phi+chart}).
\item[-] The relative-error-bounds of using co-transformation technique show similar trends
(see Figure~\ref{fig:Phi-chart}) .
\item[-] For the same interpolation technique (first-order Taylor approximation or error-correction)
and the same values of parameters ($\Delta$ and $\Delta_P$),
the relative error-bounds of using co-transformation technique to compute $\Phi^-$ in $(-1,0)$  
is between 3 to 5 times greater than those of computing $\Phi^-$ outside of $(-1,0)$ (i.e., without co-transformation) (see Figure~\ref{fig:Phi-chart}) 
\end{compactitem}
However, there is an important observation that when $\Delta = 2^{-10}$, 
the relative error-bound of using first-order Taylor approximation 
is smaller than that of using the error-correction technique with both values of $\Delta_P$ (see Appendix).
This observation can be explained 
according to the error-bound formulas of
first-order Taylor approximation (Theorem~\ref{theo1}) 
and of the error-correction technique (Theorem~\ref{the2}) as follows:
\begin{compactitem}
\item With greater values of $\Delta$,
the error-bound of first-order Taylor approximation without 
fixed-point rounding ($E_M$) is significantly greater than $\epsilon$,
so, the error-bound of the error-correction technique is dominated by the term 
$E_M(Q_S + 1- P_c(\Delta - \Delta_P) + \epsilon)$, while 
the error-bound of first-order Taylor approximation is dominated by the term $E_M$.
Intuitively, both error-bounds are dominated by their interpolation error component,
not their fixed-point rounding error component.
Because $Q_S$ and $1- P_c(\Delta - \Delta_P)$  are error-bounds of Ratio-err
and Index-err, the term $(Q_S + 1- P_c(\Delta - \Delta_P) + \epsilon)$ is small.
Therefore, the error-bound of the error-correction technique is much smaller than that of
first-order Taylor approximation.
\item In contrast, when $\Delta$ is small (approaches $\epsilon$), 
the error-bound of the error-correction technique is dominated by the term 
$(4+\Delta)\epsilon$, 
while the error-bound of first-order Taylor approximation 
is dominated by the term $(2+\Delta)\epsilon$.
Intuitively, in this case, 
both error-bounds are dominated by their fixed-point rounding error component;
and because the error-correction technique requires more 
fixed-point calculations,
its error-bound is greater than
that of first-order Taylor approximation.
\end{compactitem}
In summary, we can conclude that :
\begin{compactitem}
    \item Because the number of parameters is limited and the error-bounds computation is fast, 
    our derived error-bound formulas can effectively support manual or brute-force parameter tuning
    for LNS's desired accuracy and ROM storage.
    \item In low-precision LNS, the error-correction technique may have a counter-effect to improving accuracy.
\end{compactitem}



\section{Concluding Remarks}
\label{sec:conc}

We now briefly recap the contributions of this paper and also point out some future directions.

\subsection{Contribution}
We summarized our contribution as follows:
\begin{compactenum}
    \item We derive rigorous error-bounds for LNS's addition and subtraction, 
    which is based on combining ROM look-up tables with interpolation techniques.
    We cover the three main techniques, which are proposed in European Logarithm Microprocessor~\cite{EM}:
first-order Taylor approximation, the error-correction technique,
and the co-transformation technique.
The derived error-bound are parameterized by hardware implementation's parameters: 
$\epsilon, \Delta, \Delta_p, \Delta_a$ and $\Delta_b$.

\item We verify the correctness and tightness of our error-bounds by empirical testing.

\item We perform experimental analysis about the impact of each parameter on the magnitute of
the error-bound.
\end{compactenum}
\subsection{Future Directions}
\paragraph{Adaptations to other LNS}

For multi-based LNS designs such as~\cite{LNSbeyond2} and~\cite{madam}, 
we can simply get the error-bound for base $b$ by replacing $\log_2$ and $\ln 2$ in all 
of the error-bound formulas by $\log_b$ and $\ln b$ respectively.
For LNS designs which use different values of $\Delta$ for different
segments in $(-\infty,0)$ such as~\cite{EM},
similar to Theorem~\ref{theo1} the error-bound of first-ordered Taylor approximation of each
segment $(a,b] \subset \mathbb{R}_{\leq 0}$ is $E_{\Delta_i}(b)$ with $\Delta_i$
is the value of $\Delta$ of the segment. 
The error-bound of the whole domain $\mathbb{R}_{\leq 0}$ is simply the maximum of
all segments' error-bounds.
Deriving rigorous error-bound for ROM-less LNS is beyond the scope of this paper.
However, the formula of the error-bound of co-transformation technique is independent of 
the interpolation technique, so it is reusable to combine with any interpolation technique.


\paragraph{Mechanical Proof-Checking, Implementation Validation}
Ideally, the rigorous error-bound analysis
derived manually here
must be formally
analyzed through theorem-proving.
Given the imminent uptake of LNS, we plan to invest in some efforts with respect to theorem-proving---at least assessing the tools that best suit our purposes.
We also plan to prototype LNS on FPGAs 
or through efficient GPU kernel implementations, evaluating various tradeoffs and plugging them in lieu of other arithmetic schemes in machine learning and HPC (where appropriate).
Given that table-less methods will be more hardware efficient, the analysis of popular schemes (e.g., Mitchell's method) seem natural to pursue.

\paragraph{Pragmatic Considerations}
The advantages of LNS with respect to multiplication and division will   diminish with reduced bit sizes.
This fact, together with the overheads of addition and subtraction error correction may make other number systems (or even traditional floating-point arithmetic) the best choice for very low number sizes.
Where this cross-over occurs will be fruitful to research and discover.
Last but not least, given the wide use of dot-products in both high-performance computing and machine learning, schemes to perform dot-products and obtain reasonable error bounds will also be of immense practical interest, and in our targeted list of goals.
%

\bibliographystyle{unsrt}

\bibliography{bibfiles/ganesh,bibfiles/ganesh-nsfmed-2019,bibfiles/references}

\appendix
\section*{Appendix}
\label{sec:app1}

\renewcommand{\thesubsection}{\Alph{subsection}}

\subsection{Proof of the correctness of the co-transformation technique}
\label{ss:proofcotrans}
\begin{proof}
We prove the correctness of each case:

\noindent$\bullet\; ${\bf Case 1:\/} 
Trivial\\
\vspace{1ex}
\noindent$\bullet\; ${\bf Case 2:\/} 
From the definition of $r_b$ and $r_a$:
$$2^i-2^j = (2^i - 2^{j+r_a})-(2^j - 2^{j+r_a}) 
= 2^{i+\Phi^-(r_b)} - 2^{j+\Phi^-(r_a)} $$
The subtraction $2^i-2^j $ is transformed into the subtraction of  $2^{i+\Phi^-(r_b)} - 2^{j+\Phi^-(r_a)} $.\\
We apply the LNS subtraction's formula:
$$\log_2(2^{i+\Phi^-(r_b)} - 2^{j+\Phi^-(r_a)}) = i+\Phi^-(r_b) + \Phi^-(j+\Phi^-(r_a) - i -\Phi^-(r_b) )
= i+\Phi^-(r_b) + \Phi^-(k)$$
It is obvious that $r_a$ and $r_b$ are among the index values of $T_a$ and $T_b$,
so $\Phi^-(r_a)$ and $\Phi^-(r_b)$ are indexed directly from $T_a$ and $T_b$. 
Because $r_a \in [-\Delta_a,0)$, $\Phi^-(r_a)$ is a great negative number, which ensures
that $k <-1$ and $\Phi^-(k)$ can be computed by interpolation. 

\vspace{1ex} 
\noindent$\bullet\; ${\bf Case 3:\/} 

From the definition of $r_c$,$r_{ab}$,$r_b$ and $r_a$:
$$2^i - 2^j = (2^i-2^{j+r_{ab}}) - 
((2^{j} -2^{j+r_b})- (2^{j+r_{ab}} - 2^{j+r_b}) )$$
$$=2^{i+\Phi^-(r_c)} - ( 2^{j+\Phi^-(r_b)} - 2^{j+r_{ab}+\Phi^-(r_a)} )$$
We apply the LNS subtraction's formula: 
$$ 2^{j+\Phi^-(r_b)} - 2^{j+r_{ab}+\Phi^-(r_a)} = 
2^{j+\Phi^-(r_b) + \Phi^-(r_{ab}+\Phi^-(r_a) - \Phi^-(r_b))}
=2^{j+\Phi^-(r_b) + \Phi^-(k_1)}$$
$$\Rightarrow 2^i - 2^j  = 2^{i+\Phi^-(r_c)} - 2^{j+\Phi^-(r_b) + \Phi^-(k_1)}$$
We apply the LNS subtraction's formula one more time: 
$$\log_2 (2^{i+\Phi^-(r_c)} - 2^{j+\Phi^-(r_b) + \Phi^-(k_1)}) 
= i+\Phi^-(r_c)+\Phi^-(j+\Phi^-(r_b) + \Phi^-(k_1) - i - \Phi^-(r_c)) 
=  i+\Phi^-(r_c)+\Phi^-(k_2)$$
Similar to Case 2, $\Phi^-(r_a)$,$\Phi^-(r_b)$,$\Phi^-(r_c)$ 
are indexed directly from $T_a$, $T_b$, and $T_c$. 
Moreover, $k_1$ and $k_2$ are correspondingly dominated 
by the negative factor $\Phi^-(r_a)$ and $\Phi^-(r_b)$, 
which ensures that $k_1, k_2< - 1$,
 so $\Phi^-(k_1)$ and $\Phi^-(r_a)$ can be computed by interpolation.
\end{proof}

\subsection{Proof of Lemma~\ref{lem02}}
\label{ss:prooflem52}

\begin{proof}
Let:
$$E(\mathbf{i},\mathbf{r}) =  \hat{\Phi^-}_T(x) - \Phi^- (x)   = (\Phi^-(\mathbf{i}) - \mathbf{r}(\Phi^-)'(i)) - \Phi^-(\mathbf{i}-\mathbf{r}) $$
Similar to the proof of Lemma~\ref{lem01},
it is safe to consider the domain of $\mathbf{i}$ to be $\mathbb{R}_{\leq -1}$ 
instead of $\Delta\mathbb{Z}_{\leq-1}$  when deriving error-bound.
Then the lemma is proved by:
\begin{compactenum}
    \item Proving that $\forall \mathbf{i} \leq -1, 0 \leq \mathbf{r} < \Delta: E(\mathbf{i},\mathbf{r}) \geq 0 $
    , so $E(\mathbf{i},\mathbf{r})  = |E(\mathbf{i},\mathbf{r})| = |\Phi^+ (x)  - \hat{\Phi^+}_T(x) | $
    \item Proving that $\forall \mathbf{i} \leq -1, 0 \leq \mathbf{r} < \Delta: \frac{\partial E }{\partial \mathbf{r}}(\mathbf{i},0) \geq 0$ and $\frac{\partial E }{\partial \mathbf{i}}(\mathbf{i},0) \geq 0$, 
    so $\max_{\mathbf{i} \in \mathbb{R}_{\leq 0}, 0 \leq \mathbf{r} <\Delta} E(\mathbf{i},\mathbf{r}) <
    E(-1,\Delta) = -E^+_\Delta(-1) $
\end{compactenum}
First, we take the first and second derivatives of $E(\mathbf{i},\mathbf{r})$ w.r.t  $\mathbf{r}$:
$$ \frac{\partial E }{\partial \mathbf{r}}(\mathbf{i},\mathbf{\mathbf{r}}) = 
\frac{2^\mathbf{i}}{1 - 2^\mathbf{i}} - \frac{2^{\mathbf{i}-\mathbf{r}}}{1 - 2^{\mathbf{i}-\mathbf{r}}}     
\quad \text{and} \quad
\frac{\partial^2 E }{\partial \mathbf{r}^2}(\mathbf{i},\mathbf{\mathbf{r}}) = 
\frac{ 2^{\mathbf{i}-\mathbf{r}} \ln 2 }{(1 - 2^{\mathbf{i}-\mathbf{r}})^2} $$
From $\frac{\partial^2 E }{\partial \mathbf{r}^2} (\mathbf{i},\mathbf{r}) > 0$ and $\frac{\partial E }{\partial \mathbf{r}}(\mathbf{i},0)=0$, we conclude that $ \forall \mathbf{i} \leq -1, 0 \leq \mathbf{r} < \Delta:
 \frac{\partial E }{\partial \mathbf{r}}(\mathbf{i},\mathbf{r}) >0$.
\\
Then, because $E (\mathbf{i},0) =0$, we conclude that $ \forall \mathbf{i} \leq -1, 0 \leq \mathbf{r} < \Delta:
 E (\mathbf{i},\mathbf{\mathbf{r}}) \geq 0$.\\
We complete the proof by proving that $\forall \mathbf{i} \leq 0, 0 \leq \mathbf{r} < \Delta: 
\frac{\partial E}{\partial \mathbf{i}} (\mathbf{i},\mathbf{\mathbf{r}}) \geq 0:$
\\
When $\mathbf{i} \leq -1, 0 \leq \mathbf{r} < \Delta$, let $a= \mathbf{r}\ln 2 $
$$ \frac{\partial E}{\partial \mathbf{i}} (\mathbf{i},\mathbf{\mathbf{r}}) = 
\frac{2^\mathbf{i}}{(1-2^\mathbf{i})^2 (1-2^{\mathbf{i}-\mathbf{r}})}
(-2^\mathbf{i} f(a) + g(a))$$
$$\text{with} \quad f(a) = a e^{-a} + e^{-a} - 1 \quad \text{and} \quad
 g(a) = e^{-a} + a - 1
$$
Because $g(0) = 0$ and $g'(a) = 1 - e^{-a} \geq 0$ (for $a\geq 0$), 
we conclude that $g(a) \geq 0$.
Moreover, $\mathbf{i}-\mathbf{r} \leq 0$ and $f(a) \leq 0$ for all $a\geq 0$ as proven in Lemma~\ref{lem01}.
Hence, $\forall \mathbf{i} \leq 0, 0 \leq \mathbf{r} < \Delta: \frac{\partial E}{\partial \mathbf{i}} (\mathbf{i},\mathbf{\mathbf{r}}) \geq 0$

\end{proof}

\subsection{Proof of Lemma~\ref{lem6}}
\label{ss:prooflem66}
\begin{proof}
From~\cref{lem5}: $\forall \mathbf{i}\leq -1, \mathbf{r}\in [0, \Delta): Q^-(\mathbf{i},\mathbf{r}) \geq Q^-(-1,\mathbf{r}) = Q^-_{max}(\mathbf{r})$ and
$Q^-(\mathbf{i},\mathbf{r}) > \lim_{\mathbf{i} \to -\infty} Q^-(\mathbf{i},\mathbf{r}) = Q^-_{inf}(\mathbf{r})$ (See Lemma~\ref{lem1}). 
Therefore,
$\forall \mathbf{i}\leq 0,$ $c \leq 0 $, $r\in [0, \Delta): 
|Q^-(\mathbf{i},\mathbf{r}) - Q^-(c,\mathbf{r})|  < Q^-_{max}(\mathbf{r}) - Q^-_{inf}(\mathbf{r})$.
\\
Let $F(\mathbf{r}) = Q^-_{max}(\mathbf{r}) - Q^-_{inf}(\mathbf{r})$, 
we want to prove that $\mathbf{r}^* = \text{argmax}_{[0,\Delta)} F $.
\\
The first derivative of $F$ has the form:
$$F'(\mathbf{r}) = \frac{-X\ln 2 }
{Y(2Y-1)(X\ln X -X+1)C}
(AY^2 + BY+C)$$
where
$$Y = 2^\mathbf{r}$$
$$A = 2X\ln X -2X\ln(2X-1)+2X-2 $$
$$B = -4X\ln X +3X\ln(2X-1)-2X+2 $$
$$C = 2X\ln X -X\ln(2X-1)$$

There are at most two values of $Y$ such that $F'(\mathbf{r})=0$ 
(two solutions of $AY^2 + BY+C =0$).
The first value is $Y=1$, so the second one is $Y=\frac{-B}{A}-1=2^{\mathbf{r}^*}$.
Since $Y=2^\mathbf{r}$, there are two values of $r$ such that $F'(\mathbf{r})=0$:
$r=0$ and $r=\mathbf{r}^*$.
\\
We observe that $F(0) = F(\Delta)=0$ and according to~\cref{lem5}, 
$\forall \mathbf{i}\leq -1, \mathbf{r}\in [0, \Delta): F(\mathbf{r}) > 0$,
so $\text{argmax}_{[0,\Delta)} F \in (0,\Delta)$ since $F(\mathbf{r})$ is continuous.
Moreover, $F'(\text{argmax}_{[0,\Delta)} F)=0$ since $F(\mathbf{r})$ is differentiable. Hence, $\text{argmax}_{[0,\Delta)} F = \mathbf{r}^*.$

\end{proof}

\end{document}